\documentclass[preprint]{aastex631}
\usepackage{booktabs}
\usepackage{hyperref}
\usepackage{nameref}
\usepackage{lineno}
\usepackage{appendix}  

\hypersetup{linkcolor=red,citecolor=cyan,filecolor=blue,urlcolor=magenta}
\usepackage{subfigure}
\usepackage{color}
\shortauthors{Jiang et al.}

\begin{document}

\title{Leptohadronic Multimessenger Modeling of Two High-redshift (z $>$ 1) Neutrino Emission Blazar Candidates}

\correspondingauthor{Neng-Hui Liao; Xue Rui;  Da-Ming Wei}
\email{nhliao@gzu.edu.cn; ruixue@zjnu.edu.cn; dmwei@pmo.ac.cn}

\author{Xiong Jiang}
\affiliation{Department of Physics and Astronomy, College of Physics, Guizhou University, Guiyang 550025, China}
\affiliation{Key Laboratory of Dark Matter and Space Astronomy, Purple Mountain Observatory,
Chinese Academy of Sciences, Nanjing 210023, People’s Republic of China}
\affiliation{School of Astronomy and Space Science, University of Science and Technology of China, Hefei, Anhui 230026, People’s Republic of China}

\author[0000-0001-6614-3344]{Neng-Hui Liao}
\affiliation{Department of Physics and Astronomy, College of Physics, Guizhou University, Guiyang 550025, China}

\author[0000-0003-1721-151X]{Rui Xue}
\affiliation{Department of Physics, Zhejiang Normal University, Jinhua 321004, China}

\author[0000-0002-8966-6911]{Yi-Zhong Fan}
\affiliation{Key Laboratory of Dark Matter and Space Astronomy, Purple Mountain Observatory,
Chinese Academy of Sciences, Nanjing 210023, People’s Republic of China}
\affiliation{School of Astronomy and Space Science, University of Science and Technology of China, Hefei, Anhui 230026, People’s Republic of China}

\author[0000-0002-9758-5476]{Da-Ming Wei}
\affiliation{Key Laboratory of Dark Matter and Space Astronomy, Purple Mountain Observatory,
Chinese Academy of Sciences, Nanjing 210023, People’s Republic of China}
\affiliation{School of Astronomy and Space Science, University of Science and Technology of China, Hefei, Anhui 230026, People’s Republic of China}

\begin{abstract}
The blazars are one of the leading candidate sources of high-energy neutrinos. Recently, two blazars have been found to be temporally and spatially correlated with some IceCube high-energy neutrino events. The two blazars,  GB6 J2113+1121 and NVSS J171822+423948, are Flat Spectrum Radio Quasars (FSRQs) with redshifts greater than unity. In particular, NVSS J171822+423948 has a redshift of 2.7, which provides an important probe for studying the radiation processes of jets from active galactic nuclei in the early universe. To better understand the physical origin of the IceCube neutrinos, we adopt the one-zone leptohadronic model to fit the multimessenger emission of GB6 J2113+1121 and NVSS J171822+423948 during their $\gamma$-ray flaring periods and then calculate the high-energy neutrino detection probability. The chance of detecting a single muon neutrino from these two sources is found to be $\sim 2\%$ and $0.8\%$, respectively. Although such detection rates are not high mainly because of their high redshifts, our investigation strongly suggests that these sources are efficient PeV neutrino emitters. Our results also indicate that electromagnetic cascades produced by hadronic processes contribute significantly to X-ray and $\gamma$-ray emissions. However, high-energy $\gamma$-rays can be severely absorbed by the soft photon field from the broad-line region (BLR), which weakens the correlation between $\gamma$-rays and neutrinos, while suggesting a stronger connection between X-rays and neutrinos. We predict that IceCube will continue to detect neutrinos from FSRQs with redshifts greater than 1 in the future.

\end{abstract}

\keywords{neutrino astronomy; active galactic nuclei; Gamma-ray sources}

\section{Introduction} \label{sec:intro}

IceCube has detected some astrophysical neutrinos, marking the beginning of a new era in neutrino astronomy \citep[e.g.,][]{2013Sci...342E...1I,2015PhRvL.115h1102A,2023Sci...380.1338I}. The High-energy neutrinos observed by IceCube distribute isotropically across the sky, indicating that a significant portion of these neutrinos originates from extragalactic sources \citep{2015RPPh...78l6901A}. Additionally, the presence of high-energy neutrino emission has been detected within our Galaxy \citep{2023Sci...380.1338I}. TeV-PeV neutrinos could be produced by interactions of hadronic cosmic rays via photon-meson production($p\gamma$) or hadronuclear($pp$) processes. Some cosmic accelerators, such as the jets of active galactic nuclei (AGN), provide an ideal environment for the acceleration of ultra-high-energy cosmic rays (UHECRs) \citep{2012ApJ...749...63M}, and interactions of hadronic cosmic rays with the radiation fields and matter close to the source lead to the production of high-energy pions, which then decay into neutrinos. Unlike cosmic rays, however, neutrinos are not deflected by magnetic fields during their travel and interact very little with matter. Therefore, detecting TeV-PeV neutrinos is an effective method for studying extreme cosmic accelerators \citep{2018AdSpR..62.2902A}. Various extragalactic sources are suggested as potential high-energy neutrino emitters, such as galaxy clusters \citep{2008ApJ...689L.105M}, jets of AGN \citep[e.g.,][]{2005APh....23..537H,2014JHEAp...3...29D,2019NatAs...3...88G,2024arXiv241107827C}, $\gamma$-ray bursts \citep[e.g.,][]{1995PhRvL..75..386W, 2015NatCo...6.6783B}, starburst galaxies \citep{2006JCAP...05..003L}, and tidal disruption events  \citep[e.g.,][]{2022PhRvL.128v1101R,2023ApJ...953L..12J}. Additionally, the Seyfert galaxy NGC 1068 has been found to be associated with high-energy neutrino emission with a global significance of 4.2$\sigma$ \citep{2022Sci...378..538I}, but no firm  connection has been established between NGC 1068 and neutrino detections.
Due to the relatively large positional uncertainty of IceCube neutrinos, additional data are needed to robustly infer the sources of extragalactic astrophysical neutrinos. 
Hadronic cosmic rays interactions not only produce neutrinos but also lead to the emission of $\gamma-$rays, which come from neutral pion decays and cascade emissions of secondary electron-positron pairs, as well as X-rays generated by the synchrotron radiation from these secondary pairs. A $\gamma$-ray flare is often caused by an increase in the luminosity of relativistic electrons in the source, an enhanced apparent Doppler factor, and/or the strengthening of the target photon field. Assuming that protons are co-accelerated with electrons in the jet, these factors can significantly enhance neutrino production efficiency \citep[e.g.,][]{2014PhRvD..90b3007M,2016APh....80..115P,2018ApJ...854...54R}. Additionally, the decay of neutral pions and cascade radiation produced by hadronic processes can further enhance $\gamma-$rays emission. However, $\gamma$-rays may be absorbed in a dense low-energy radiation field, such as ultraviolet or X-rays, which serves as an efficient target for neutrino production through the $p\gamma$ process. As a result, it remains unclear whether neutrinos are associated with $\gamma$-ray flares. If the $\gamma$-rays produced alongside neutrinos are able to escape from the sources, simultaneous observations of neutrino events and $\gamma$-ray flares would provide an effective way to study the origins of neutrinos. A notable case is the detection of GeV and TeV $\gamma$-ray flares, associated with a high-energy neutrino event, from the blazar TXS 0506+056 \citep{2018Sci...361.1378I}.

Blazars, a type of AGN characterized by relativistic jets oriented toward Earth \citep{2019ARA&A..57..467B}, have been proposed as potential sources of UHECRs and/or high-energy neutrinos \citep[e.g.,][]{1993A&A...269...67M,1997ApJ...488..669H,2014MNRAS.443..474P,2016PhRvL.116g1101M,2018ApJ...865..124M}. Blazars are radio-loud AGNs characterized by a significantly broader energy spectrum that exhibits Doppler-boosted non-thermal emission across all wavelengths, from radio up to TeV $\gamma-$ray, along with rapid variability and a high degree of polarization in the radio and optical wavebands \citep{2016ARA&A..54..725M}. Based on their optical spectrum features, blazars can be classified into  Flat Spectrum Radio Quasars (FSRQs) and BL Lacertae objects (BL Lacs). FSRQs have strong broad emission lines in their optical spectrum  \citep{2012ApJ...748...49S}, indicating the presence of a radiatively efficient accretion disk and a BLR composed of gas clouds that intercept a portion of the radiation emitted by the accretion disk and re-radiate it in the form of emission lines \citep{2009MNRAS.396L.105G}. While BL Lacs  have very weak emission lines or their optical spectrum is featureless \citep{2014A&ARv..22...73F},  which indicates that the majority of them possess a significantly less radiatively efficient accretion disk. 

The spectral energy distribution (SED) of a blazar typically exhibits two prominent bumps: one that peaks at low energies (infrared to X-rays), usually attributed to synchrotron emission from a population of relativistic electrons in the jet, and another that peaks in the $\gamma$-ray range \citep{1996MNRAS.280...67G}. Based on the location of their synchrotron emission peak frequencies ($\nu^S_{peak}$), blazars can be further classified into low-synchrotron peaked objects (LSP; $\nu^S_{peak} < 10^{14}$ Hz  ), intermediate-synchrotron peaked objects (ISP; $10^{14}$ Hz  $< \nu^S_{peak} < 10^{15}$ Hz  ), and high-synchrotron peaked objects (HSP; $\nu^S_{peak} > 10^{15}$ Hz  ) \citep{2010ApJ...716...30A}. The origin of the high-energy emission from blazars is still a subject of ongoing debate. In leptonic models, it is assumed that the high-energy emission  results from low-energy soft photons undergoing inverse-Compton (IC) scattering with the relativistic electrons in the blazar jet. If the soft photons originate from the synchrotron radiation of relativistic electrons, it is referred to as synchrotron-self-Compton (SSC) emission, which is commonly used to explain the $\gamma$-ray emissions of BL Lacs \citep{1996ApJ...461..657B}. Conversely, if the soft photons come from external sources, it is referred to as external-Compton (EC) emission, which is typically used to explain the $\gamma$-ray emissions of FSRQs \citep{1994ApJ...421..153S}. In hadronic models, the high-energy bump of blazars can arise from protons synchrotron radiation \citep[e.g.,][]{2000NewA....5..377A, 2001APh....15..121M} or from emission by secondary pairs generated in the $p\gamma$ interaction \citep[e.g.,][]{1993A&A...269...67M, 2003APh....18..593M, 2003ApJ...586...79A}.

Blazars are the most numerous $\gamma$-ray sources in the extragalactic sky, and considering that the production of neutrinos inevitably accompanies the emission of $\gamma$-ray, this is one of the reasons they have long been considered primary candidates for neutrino sources. Several models have been developed to explain the electromagnetic radiation or neutrino emissions of blazars, including aspects of time evolution and spectral behavior, such as the one-zone leptohadronic model \citep[e.g.,][]{2012A&A...546A.120D,2013ApJ...768...54B,2015MNRAS.448..910C,2022MNRAS.509.2102G, 2023arXiv231213371K, 2024A&A...683A.225S}, the multi-zone  leptohadronic model \citep[e.g.,][]{2019ApJ...886...23X,2021ApJ...906...51X}, and the hadronuclear model \citep[e.g.,][]{2019PhRvD..99f3008L, 2020PhRvD.101f3024B, 2021RAA....21..305W, 2024ApJ...971..146X}.

On September 22, 2017, IceCube detected a neutrino with an energy of about 0.3 PeV, named IceCube -170922A. Within the positional uncertainty of this neutrino, the blazar TXS 0506+056 \citep[$z$ = 0.3365;][]{2018ApJ...854L..32P} was experiencing a 6-month-long $\gamma$-ray flare \citep{2018Sci...361.1378I}. The connection between IceCube-170922A and the $\gamma$-ray flare from TXS 0506+056 was found with a confidence level of about 3$\sigma$ . The observed neutrino and $\gamma-$ray emissions have been interpreted using a one-zone hybrid leptohadronic model with $p\gamma$ interaction \citep[e.g.,][]{2018ApJ...863L..10A, 2018ApJ...864...84K, 2019NatAs...3...88G, 2019MNRAS.483L..12C, 2019MNRAS.489.4347O}. Additionally, a search for further neutrinos from TXS 0506+056 in IceCube data uncovered evidence of a neutrino flare during 2014–2015 with a significance of  3.5$\sigma$, but no corresponding electromagnetic activity was observed \citep{2018Sci...361..147I}. Nevertheless, this further supports the notion that the blazar TXS 0506+056 is a neutrino-emitting source. 
TXS 0506+056 is classified as a BL Lac object because its optical spectrum shows an apparent absence of broad emission lines, and, like TXS 0506+056, an increasing number of BL Lac objects are being found to have temporal and spatial correlations with individual high-energy neutrino events \cite[e.g.,][]{2019ApJ...880..103G,2020ApJ...893..162F,2020A&A...640L...4G,2023MNRAS.519.1396S}. In addition to BL Lacs, FSRQs have also been observed to have temporal and spatial correlations with neutrinos. A notable example is the high-luminosity FSRQ PKS 1424-418, for which a temporal and directional coincidence was observed between its $\gamma-$ray outburst and a PeV-energy neutrino event \citep{2016NatPh..12..807K}. However, none of these associations are significant.

Recently, \cite{2022ApJ...932L..25L} reported a temporal and spatial correlation between the neutrino IceCube-191001A and an electromagnetic flare from the blazar GB6 J2113+1121 at a redshift of 1.3. About half a year before the arrival of IceCube-191001A, GB6 J2113+1121 underwent a $\gamma$-ray flare that was unprecedented since the start of the Fermi Large Area Telescope (Fermi-LAT) operation. This flare lasted for $\sim$ 1 year (see Figure \ref{fA.1}  in the appendix). Simultaneously, violent optical flares were detected in the ZTF $g$, $r$, and $i$ bands, along with infrared flares in the WISE W1 and W2 bands (see Figure \ref{fA.2} in the appendix).
Additionally, \cite{2024ApJ...965L...2J} found that the neutrino IceCube-201221A also exhibits a temporal and spatial correlation with an electromagnetic flare from the blazar NVSS J171822+423948 at a redshift of 2.7. NVSS J171822+423948 remained in a quiescent state for $\sim$ 12 years, undetected by Fermi-LAT, but shortly after the arrival of the neutrino (within a few tens of days), its $\gamma$-ray flux increased tenfold and persisted for $\sim$ 2 years (see Figure \ref{fA.3} and \ref{fA.4} in the appendix). Meanwhile, its infrared flux also began to rise. Furthermore, 115 days after the neutrino’s arrival, strong optical flares were observed in the ZTF $g$, $r$, and $i$ bands. Considering that the $\gamma$-ray activity of both sources is correlated with the neutrinos at $\sim$ 2.2$\sigma$, this suggests that the FSRQs GB6 J2113+1121 and NVSS J171822+423948 are likely among the sources of neutrinos detected by IceCube. Unlike TXS 0506+056, which has a redshift of 0.335, both GB6 J2113+1121 and  NVSS J171822+423948 are high-redshift blazars, with redshifts greater than 1. Notably,  NVSS J171822+423948 has a redshift of 2.7, making it the most distant blazar known to be temporally and spatially correlated with neutrinos. 

Only the most luminous high-redshift blazars ($z \geq 2$) can be detected in $\gamma$-ray emission, and these luminous sources are the most powerful persistent $\gamma$-ray sources in the universe, exhibiting the highest jet power and luminosity, with central black hole masses typically exceeding $10^9$ solar masses. Studying these blazars helps to further understand the cosmological evolution of blazars and supermassive black holes, as well as the evolution of relativistic jets across different cosmic epochs \citep{2024ApJ...974...38G}. In the early universe, high-redshift blazars are crucial for investigating relativistic jets and their connection with the central engine, namely the black hole and the accretion disk. Studies have shown that supermassive black holes in jetted AGNs evolve faster than those in radio-quiet \citep{2015MNRAS.446.2483S}, suggesting a link between jet activity and black hole growth \citep{2014MNRAS.442L..81F}. Due to their great distance, the high-energy emission of these objects typically peaks below the GeV range, making them difficult to study with the Fermi-LAT. Moreover, most of these high-redshift sources can only be detected by the LAT during flaring states \citep{2019ApJ...871..211P}. As a result,  to date, only a small number of high-redshift blazars have been detected by Fermi-LAT. The association between neutrinos and flares of high-redshift blazars is extremely rare, while the study of multi-messenger emissions from individual objects provides valuable insights into the general physics of high redshift blazars.  Therefore, we are motivated to explore the relationship between electromagnetic emissions and neutrinos in these two FSRQs by conducting detailed multimessenger modeling during their $\gamma$-ray flares.

This paper is structured as follows: In Section \ref{sec:2}, we introduce the general description of the one-zone leptohadronic model used for the calculation of the electromagnetic  emission and neutrino emission of GB6 J2113+1121 and NVSS J171822+423948. In Section \ref{sec:3}, we present the relevant results of the SED modeling and provide a corresponding report on them. In Section \ref{sec:4}, we discuss the impact of different soft photon fields on neutrino production and compare GB6 J2113+1121 and NVSS J171822+423948 with other neutrino candidates. In Section \ref{sec:5}, we provide a concise summary of our findings to conclude the paper. We adopt a $\Lambda$CDM cosmology with $ \Omega_{M} $ = 0.32, $ \Omega_{\Lambda} $ = 0.68, and a Hubble constant of $H_{0}$ = 67 km$^{-1}$ s$^{-1}$ Mpc$^{-1}$ \citep{2014A&A...571A..16P}.

\section{Model Description} \label{sec:2}

In our modeling, we consider a simple one-zone  leptohadronic model scenario to explain the electromagnetic and neutrino radiation, assuming that both the electromagnetic and neutrino emissions originate from the same radiation region.
We assume that the region responsible for blazar emission can be characterized as a spherical blob, moving with a bulk Lorentz factor $\Gamma$ relative to the central black hole. The Doppler factor of the jet is given by   $\delta = \Gamma^{-1}(1-\beta{cos\theta})^{-1}$, where $\theta$ denotes the angle between the line of sight and the jet axis, and $\beta = \vartheta/c$, with $c$ being the speed of light and $\vartheta$ being the speed of the jet. In our work, we
assume the jet is observed at an angle $\theta \sim 1/\Gamma$ relative to its axis, therefore $\delta \sim \Gamma$. The blob, filled with relativistic electrons and protons, along with an isotropic magnetic field, allows primary electrons to produce synchrotron radiation under the influence of the magnetic field, which explains the low-energy part of the SED. The high-energy part of the SED can be explained by inverse Compton scattering of soft photons by the primary electrons or proton-induced cascade emission. Here, we do not consider proton synchrotron emission as an explanation for the $\gamma-$rays, because this scenario requires rather extreme physical parameters, such as magnetic fields typically reaching several tens to hundreds of Gauss, and it fails to simultaneously reproduce the electromagnetic SED and the energy and flux of the neutrino emission observed by IceCube \citep{2019NatAs...3...88G, 2019MNRAS.483L..12C, 2022MNRAS.509.2102G, 2023MNRAS.519.1396S}. In blazar hadronic models, $pp$ interactions are often overlooked, as interactions of relativistic protons with ambient matter (gas) require high gas densities to explain the observed luminosities \citep{2002A&A...382..829S,2011A&A...531A..30R}. Moreover, \cite{2024ApJ...971..146X} have suggested that $pp$ interactions are only important when the  emission region of low-redshift sources is located at the base of the jet. Meanwhile, the density of the soft photon field from the BLR or the co-accelerated electrons typically exceeds that of the gas. Therefore, we naturally consider that neutrinos originate from the decay of pions produced in the $p\gamma$ process (see section \ref{subsec:2-2}). Secondary pairs generated in the $p\gamma$ process and Bethe-Heitler pair production process trigger electromagnetic cascades, which, under the influence of the magnetic field, mainly contribute to the electromagnetic radiation through synchrotron emission.

The specific description of our model is as follows. For simplicity, quantities with the superscript "obs" refer to the observer frame, quantities with a prime superscript are measured in the cosmic rest frame, and quantities without a superscript refer to the jet comoving frame, unless otherwise specified.
\subsection{Leptonic processes} \label{subsec:2-1}

We consider that relativistic electrons are injected into the radiation region at a constant rate, with their energy distribution following a broken power law with an exponential cut-off:
\begin{equation}
Q_e(\gamma_e) = 
\left\{
\begin{array}{ll}
Q_{e,0} \gamma_e^{-n_{e,1}}  & \text{for } \gamma_{e,\text{min}} \leq \gamma_e < \gamma_{e,\text{break}} \\
\gamma_{e,\text{break}}^{n_{e,2} - n_{e,1}} Q_{e,0} \gamma_e^{-n_{e,2}} e^{-\gamma_e/\gamma_{e,max}} & \text{for } \gamma_{e,\text{break}} \leq \gamma_e \leq \gamma_{e,max}
\end{array}
\right. ,
\end{equation}
where $\gamma_{e,min}$, $\gamma_{e,break}$, and $\gamma_{e,max}$ are the minimum, break, and maximum electron Lorentz factors, respectively; $n_{e,1}$ and $n_{e,2}$ are the power-law indices below and above $\gamma_{e,break}$, respectively; and $Q_{e,0}$ is the normalization factor of the electron distribution, in units of {cm}$^{-3}$ {s}$^{-1}$. The  normalization factor $Q_{e,0}$ can be obtained from  $\int Q_e \gamma_e m_e c^2 d\gamma_e = 3 L_{e,\text{inj}}/(4 \pi R^3)$. Here, $m_e$ is the electron rest mass, $c$ is the speed of light, $L_{e,inj}$ is the electron injection luminosity, and $R$ is the radius of the  blob. Then, the inject electrons cool via synchrotron emission and inverse Compton scattering. To obtain a steady-state radiation, we need to determine the steady-state electron distribution $N_e(\gamma_e)$, which requires solving the time-dependent transport equation for electrons. The time-dependent transport equation is described as :
\begin{equation}
\frac{\partial}{\partial t} N_e(\gamma_e) = \frac{\partial}{\partial \gamma_e} \left[ \gamma_e \frac{N_e(\gamma_e)}{\tau_c(\gamma_e)} \right] + Q_e(\gamma_e) - \frac{N_e(\gamma_e)}{\tau_{\text{ad}}}   ,
\label{eq2}
\end{equation}
where $\tau_{ad}$ = $R/c$ is the energy-independent escape time, and 
\begin{equation}
\tau_c(\gamma_e) = \frac{3m_ec}{4(u_B + u_{soft}(\gamma_e))\sigma_T}\frac{1}{\gamma_e}  ,
\end{equation}represents the radiative cooling time, which incorporates both synchrotron losses(with $u_B$ = $B^2/(8\pi)$ as the magnetic energy density) and inverse Compton losses(in the Thomson regime, where $u_{soft}(\gamma_e)$ is the soft photon energy density, $\sigma_T$ is the Thomson scattering
cross-section).The $u_{soft}(\gamma_e)$ can be described as \citep{1995A&A...295..613M} 
\begin{equation}
u_{soft}(\gamma_e) = m_ec^2 \int_0^{\frac{3}{4\gamma_e}} \epsilon_2 n_{ph}(\epsilon_2) \, d\epsilon_2   ,
\end{equation}
where $n_{ph}(\epsilon_2)$ is the number density distribution of soft photons in the jet comoving frame.
We can get the steady-state electron distribution according to the formula provided by \cite{1996ApJ...463..555I}

\begin{equation}
N_e(\gamma_e) = e^{-\gamma_e^*/\gamma_e} \frac{\gamma_e^*\tau_{ad}}{\gamma_e^2} \int_{\gamma_e}^{\infty} Q_e(\psi)e^{+\gamma_e^*/\psi} \, d\psi  ,
\label{eq5}
\end{equation}

where
\begin{equation}
\gamma_e^* = \frac{3m_ec^2}{4(u_B + u_{soft}(\gamma_e))\sigma_TR}  ,
\end{equation}
is the cooling break Lorentz factor when $\tau_c(\gamma_e)$ = $\tau_{ad}$.  

Then we can compute the associated stationary synchrotron emission \citep{2009herb.book.....D}, which includes synchrotron self-absorption, as well as IC scattering \citep{1970RvMP...42..237B} using the steady-state electron distribution $N_e(\gamma_e)$.
 The soft photon field for SSC emission originates from the synchrotron radiation of the primary electrons. 
For EC emission, the energy density of BLR radiation in the jet comoving frame ($u_{BLR}$) can be described as a function of the distance along the jet \citep{2012ApJ...754..114H}
\begin{equation}
u_{BLR} = \frac{\Gamma^2L^{'}_{BLR}} 
{4\pi{(r_{BLR}^{'})}^2c[1+{(r_{in}^{'}/r_{BLR}^{'})}^3]}  ,
\end{equation}
where  $r_{BLR}^{'}$ = 0.1$(L_{BLR}^{'}/{10}^{45} erg s^{-1})^{1/2}$ pc is the characteristic radius of the BLR \citep{2008MNRAS.387.1669G}, $r_{in}^{'}$  is the distance from the central black hole to the radiation region, and $L^{'}_{BLR}$ denotes the BLR radiation luminosity, given by: 
\begin{equation}
L^{'}_{BLR} = f_{BLR}^{'}L_{disk}^{'} ,
\end{equation}
where $L_{disk}^{'}$ is the accretion disk luminosity, and $f_{BLR}^{'} = 0.1$ is the fraction of the disk luminosity reprocessed into the BLR radiation \citep{2008MNRAS.387.1669G}.
The BLR emission spectrum is composed of a continuum and Doppler-broadened lines, with the Ly$\alpha$ line being the most significant contribution. The spectrum of NVSS J171822+423948 exhibits a Ly$\alpha$ emission line,  while the spectrum of GB6 J2113+1121 shows a broad $C_\mathrm{IV}$ emission line. The energy spectrum of the BLR emission can be well approximated by a single-temperature blackbody, peaking at $\sim 2 \times {10}^{15}\Gamma$ Hz \citep{2008MNRAS.386..945T}, as measured in the jet comoving frame. In our modeling, we did not consider the emission lines from the BLR. Instead, for simplicity, we approximate the BLR radiation of both sources using the blackbody continuum described here. The results are consistent with those from considering  the emission lines, so the results of our study are not affected by this distinction. 
In addition to the BLR, the dusty torus surrounding the accretion disk can also reprocess the disk radiation, thereby generating thermal infrared emission. The radiation from the dust torus can be modeled as an isotropic blackbody spectrum, with a peak frequency of $3 \times 10^{13} \Gamma$ Hz in the jet comoving frame \citep{2007ApJ...660..117C}. The energy density of the dust torus radiation in the jet comoving frame can be expressed as \citep{2017MNRAS.469..255G}
\begin{equation}
u_{DT} = \frac{\Gamma^2 0.07 f_{DT}^{'}}{12 \pi} \; \text{erg} \, \text{cm}^{-3}
\end{equation}
where $f_{DT}^{'} = 0.1$ is the fraction of the disk luminosity reprocessed into dust torus radiation \citep{2012ApJ...754..114H}. 
However, in our modeling of the EC emission, we only considered the photon field from the BLR and did not take into account the photon field from the accretion disk and dust torus, because in leptonic models, it is typically assumed that the EC process on BLR photons is the primary mechanism powering the high-energy emission of most FSRQs \citep{1998MNRAS.301..451G,2009ApJ...704...38S}. On one hand, the photon field from the accretion disk becomes relevant only under stringent conditions, where the emission region lies very close to the black hole, as its energy density decreases with the square of the distance to the black hole.  On the other hand, when considering the BLR photon field, due to the larger radius of the dust torus ($r_{DT}^{'}$ = 2.5$(f_{DT}^{'}L_{disk}^{'}/{10}^{45} \text{erg}$  s$^{-1})^{1/2}$ pc), the corresponding photon energy density from the dust torus is negligible compared to the BLR emission in our model. The photon energy densities of the BLR and the dust torus in the AGN frame are $u^{'}_{BLR} \sim 3 \times 10 ^{-2} $ erg cm$^{-3}$ and  $u^{'}_{DT} \sim 2 \times 10 ^{-4} $ erg cm$^{-3}$ , respectively.

The high-energy photons produced by SSC and EC emissions will be absorbed by the soft photon field, leading to $\gamma\gamma$ annihilation. The the optical depth of a high-energy photon with energy $\epsilon_{1}$ can be described as
\begin{equation}
\tau_{\gamma\gamma}(\epsilon_{1}) = r_b \int_0^\infty n_{ph}(\epsilon_2) \sigma_{\gamma\gamma}(s) \, d\epsilon_2  ,
\end{equation}
where $r_b$ is  the absorption path length, $\sigma_{\gamma\gamma}(s)$ is the cross-section of $\gamma\gamma$ annihilation, with $s = \epsilon_{1} \epsilon_2$, $\epsilon_1 = E_1/(m_ec^2)$  and $\epsilon_2 = E_2/(m_ec^2)$ represent the normalized, dimensionless energies of the high-energy and low-energy photons, respectively. The cross-section of $\gamma\gamma$ annihilation is given by \cite{1990MNRAS.245..453C}
\begin{equation}
\sigma_{\gamma\gamma}(s) = \frac{3\sigma_T}{16}(1-s^2)[2s(s^2-2)+(3-s^4)\ln(\frac{1+s}{1-s})] ,
\end{equation}
The luminosity of the $\gamma$-ray with energy $\epsilon_\gamma$ that escapes from the source can be expressed as
\begin{equation}
L_{\gamma, escape}(\epsilon_\gamma) = L_\gamma(\epsilon_\gamma) \times \frac{1-e^{-\tau_{\gamma\gamma}(\epsilon_\gamma)}}{\tau_{\gamma\gamma}(\epsilon_\gamma)}  ,
\end{equation}
where $L_\gamma(\epsilon_\gamma)$ is the luminosity of the $\gamma-$ray that have not been absorbed. Similarly, the very high energy (VHE) $\gamma$-ray photons that escape from the source will be absorbed by the extragalactic background light (EBL). We use the EBL model of \cite{2011MNRAS.410.2556D} to compute the $\gamma$-ray attenuation, and the observed $\gamma$-ray luminosity is modified by a factor of $e^{-\tau_{\gamma\gamma}}$ after absorption by the EBL.

\subsection{ Hadronic processes}
\label{subsec:2-2}
The protons within the blob can interact with the soft photons, resulting in various photohadronic processes. 
One such process is the $p\gamma$ process, in which protons interact with soft photons to produce pions
\begin{equation}
p + \gamma \rightarrow N + \pi , 
\end{equation}
where $N$ represents proton or neutron, and $\pi$ represents $\pi^0$, $\pi^+$, or $\pi^-$. Pions subsequently decay as follows:
\begin{equation}
\begin{array}{ll}
\pi^0 \rightarrow 2\gamma ,\\
\pi^+ \rightarrow \mu^+ + \nu_{\mu} \rightarrow e^+ + \nu_e + \bar{\nu_{\mu}} + \nu_{\mu} ,\\
\pi^- \rightarrow \mu^- + \bar{\nu_{\mu}} \rightarrow e^- + \bar{\nu_e} + \nu_{\mu} + \bar{\nu_{\mu}} ,
\end{array}
\end{equation}
The other process is Bethe-Heitler pair production, in which protons interact with soft photons to generate electron-positron pairs
\begin{equation}
p + \gamma  \rightarrow p^{'} + e^+ + e^-  ,
\end{equation}

We consider that relativistic protons are injected into the radiation region blob at a constant rate, with their energy distribution following a single power law with an exponential cut-off:
\begin{equation}
Q_p(\gamma_p) = 
Q_{p,0} \gamma_p^{-n_{p}} e^{-\gamma_p/\gamma_{p.max}}  \\
\text{ }\text{for} \text{ }\gamma_{p,\text{min}} \leq \gamma_p \leq \gamma_{p,max} ,
\end{equation}
where $Q_{p,0}$ is the normalization number density which can be determined from $\int Q_p \gamma_p m_p c^2 d\gamma_p = 3 L_{p,\text{inj}}(4 \pi R^3)$ , $\gamma_{p,min}$ and $\gamma_{p,max}$ are the minimum,  and maximum proton Lorentz factors, respectively; $n_{p}$ is the power-law index.

To determine whether protons cool effectively, we compared all relevant timescales. The proton-synchrotron cooling timescale is given by $t_{p,syn} = 6\pi m_e c^2 /(c\sigma_T B^2 \gamma_p) (m_p/m_e)^3$.The $p\gamma$ energy-loss timescale is written as \citep{1997PhRvL..78.2292W}:
\begin{equation}
t_{p\gamma}^{-1} = \frac{c}{2\gamma_p^2} \int_{\bar{\epsilon}_{th}}^\infty d\bar{\epsilon} \sigma_{p\gamma}(\bar{\epsilon})k_{p\gamma}(\bar{\epsilon}) \bar{\epsilon} 
\int_{\bar{\epsilon}/2\gamma_p }^\infty d\epsilon n(\epsilon_2) \epsilon_2^{-2} ,
\end{equation}
where quantities with a bar are measured in the proton's rest frame, $\bar{\epsilon}_{th}$ = 145 MeV, $\sigma_{p\gamma}(\bar{\epsilon})$ and $k_{p\gamma}(\bar{\epsilon})$ represent the cross-section and proton inelasticity as functions of photon energy $\bar{\epsilon}$ respectively \citep{2000CoPhC.124..290M, 1968PhRvL..21.1016S}, and $n(\epsilon_2)$  is the target photon number density in the jet-comoving frame. 
The BH pair-production cooling timescale is given by \citep{1970RvMP...42..237B, 2015MNRAS.447...36P}
\begin{equation}
t_{BH}^{-1} = \frac{3}{8\pi \gamma_p} \sigma_Tc\alpha_f \frac{m_e}{m_p} \int_{2}^\infty d\kappa n_{\gamma}(\frac{\kappa}{2\gamma_p}) \frac{\phi(\kappa)}{\kappa^2}, 
\end{equation}
where $\alpha_f \approx 1/137$ is the fine structure constant, $\kappa$ = $2\gamma_p\epsilon/m_ec^2$, and $\phi(\kappa)$ is a function from equation (3.12) of \citep{1992ApJ...400..181C}.
Taking into account the proton cooling times discussed above, the steady-state proton distribution can be approximated as:
\begin{equation}
N_p(\gamma_p) = 
Q_{p,0} \gamma_p^{-n_{p}}t_p  ,
\end{equation}
where $t_p$ = min$[t_{p,syn},t_{dyn},t_{p\gamma}, t_{BH}]$ with $t_{dyn} = R/c$ being the energy-independent particles escape timescale of the blob. Once the steady-state proton distribution $N_{p}$ is obtained, we can then use the methods developed by \cite{2008PhRvD..78c4013K} to calculate the spectra of the products from the $p\gamma$ and Bethe-Heitler processes.

Photons from $\pi^0$ decay interact with low-energy photons to undergo $\gamma\gamma$ annihilation process. The attenuated $\gamma$-rays with energy $\epsilon_\gamma$ produce electron-positron pairs, with one of the produced particles assuming the major fraction $f_{\gamma_e}$ of the photon energy. Thus, the resulting electron/positron pair has energies of $\gamma_1 = f_{\gamma_e}\epsilon_\gamma$ and $\gamma_2 = (1-f_{\gamma_e})\epsilon_\gamma$, respectively. Considering that each photon that does not escape the blob will produce an electron-positron pair, the pair production rate can be written as
\begin{equation}
\dot{N}_{e}^{\gamma\gamma}(\gamma_e) = (1 - \frac{1-e^{- \tau_{\gamma\gamma}(\epsilon_{\gamma,1})}}{\tau_{\gamma\gamma}(\epsilon_{\gamma,1})}) \dot{N}^0_{\epsilon_{\gamma,1}} + (1 - \frac{1-e^{- \tau_{\gamma\gamma}(\epsilon_{\gamma,2})}}{\tau_{\gamma\gamma}(\epsilon_{\gamma,2})}) \dot{N}^0_{\epsilon_{\gamma,2}} 
,\end{equation}
where $\epsilon_{\gamma,1} = \gamma_e/f_{\gamma_e}$, $\epsilon_{\gamma,2} = \gamma_e/(1-f_{\gamma_e})$, and $f_{\gamma_e}$ = 0.9 \citep{2013ApJ...768...54B}.

$\dot{N}_{e}^{\gamma\gamma}(\gamma_e)$, along with the electrons/positrons produced by $\pi^{\pm}$ decay($Q_{e,p\gamma}$) and the electron-positron pairs generated by Bethe–Heitler pair production($Q_{e,BH}$) can be used as injection terms in equation (\ref{eq2}). The steady-state distribution of the first-generation secondary electron-positron pairs can then be obtained using equation (\ref{eq5}). However, due to the very high energy of the first-generation steady-state secondary electron-positron pairs, the IC emission is severely suppressed by the Klein–Nishina (KN) effect, while the photons produced by synchrotron emission still have very high energy. Similar to photons from $\pi^0$ decay,  these high-energy photons annihilate with soft photons to produce second-generation electron-positron pairs, which in turn can generate a third generation, and so on. To obtain the final steady-state cascade emission, we adopted the method proposed by \cite{2015MNRAS.448..910C}: We repeat this process until the contribution of electromagnetic radiation from the $i$-th generation of pairs becomes negligible compared to the sum of the contributions from the previous generations. In calculating the electromagnetic cascade, due to the KN suppression, photons from high-energy pairs do not significantly enhance the IC cooling of high-energy pairs. Therefore, we assume that the low-energy photon field originates exclusively from the synchrotron emission of primary electrons and radiation from the BLR, neglecting the electromagnetic emission from the cascade pairs themselves. This assumption has been validated by \cite{2015MNRAS.448..910C}.

\subsection{Parameter Constraints}
\label{subsec:2-3}
In our model, there are a total of 15 free parameters: 6 for  the injected primary relativistic electrons($\gamma_{e,min}$, $\gamma_{e,break}$, $\gamma_{e,max}$, $n_{e,1}$, $n_{e,2}$ and $L_{e,inj}$) , 4 for the injected primary relativistic protons($\gamma_{p,min}$, $\gamma_{p,max}$, $n_{p}$, $L_{p,inj}$), and 5 for  the emitting region($\delta$, B, R, $r_{in}^{'}$ and $L_{BLR}^{'}$).
We decrease the number of free parameters using the following methods.

1. According to  \cite{2022ApJ...932L..25L} and \cite{2024ApJ...965L...2J}, the accretion disk luminosities
$L_{disk}^{'}$ of GB6 J2113+1121  and  NVSS J171822+423948  are $\sim$ $1\times10^{46} $ erg s$^{-1}$ and $\sim$ $3\times10^{46} $ erg s$^{-1}$, respectively. Considering that a fraction of the disk luminosity is reprocessed into the BLR and using a standard covering factor of $\sim$ 0.1, the BLR radiation luminosities $L^{'}_{BLR}$ of GB6 J2113+1121  and  NVSS J171822+423948  as $\sim$ $10^{45}$ erg s$^{-1}$ and $\sim$ $3\times10^{45}$ erg s$^{-1}$, respectively. We then determined that the characteristic  radius of the BLR  $r^{'}_{BLR}$ for GB6 J2113+1121  and  NVSS J171822+423948  are $\sim$ 0.1 pc and $\sim$ 0.2 pc, respectively.

2.We set the spectral index of the injected protons to be $ n_p = 2$ as expected from first-order Fermi acceleration \citep{2007Ap&SS.309..119R}.

3. We fix the minimum Lorentz factor of protons $\gamma_{p,min}$ to be 1, because as long as its value is not too large, its impact on our fitting results is minimal. It primarily affects the total proton energy density.

4. In leptonic models, the minimum electron Lorentz factor $\gamma_{e,min}$ is typically constrained by hard X-ray data, as this energy band is where the low-energy tail of the SSC emission appears. However, in hadronic models, synchrotron emission from pair cascades makes a significant contribution to the hard X-ray band. In addition, there are no available simultaneous radio data to constrain $\gamma_{e,min}$,  so we are unable to determine it through SED fitting. In our modeling, we adopt a typical value of  $\gamma_{e,min}$ = 50 \citep{2019ApJ...886...23X}. 

\begin{figure*}
    \centering
    \includegraphics[scale=0.4]{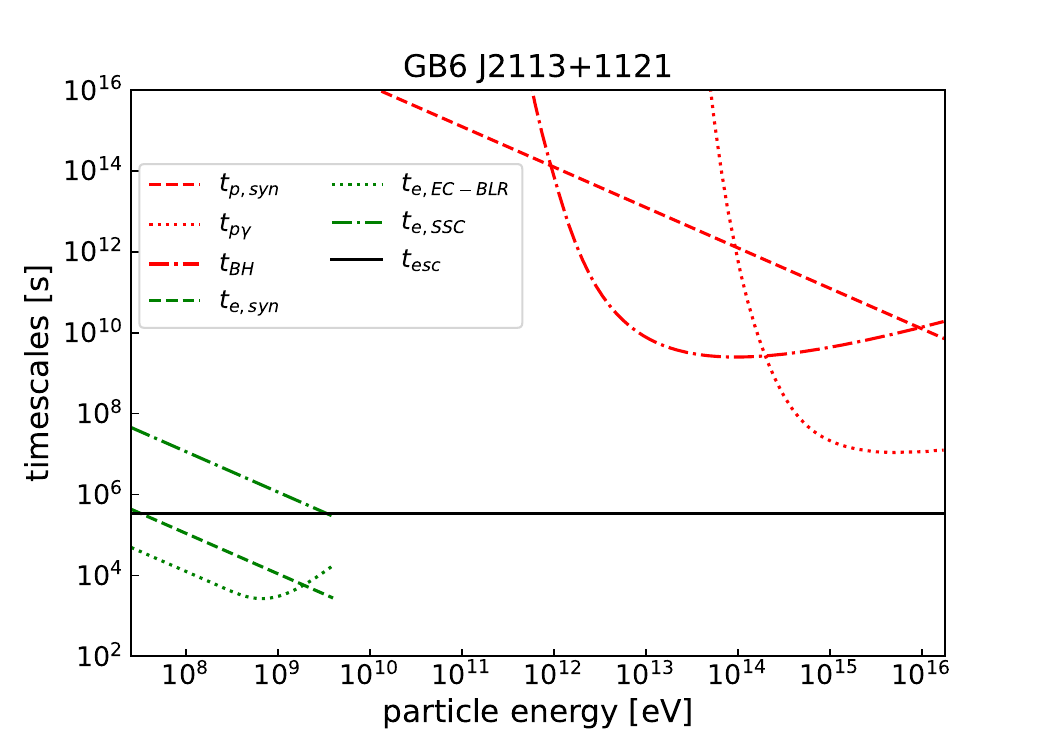}
    \includegraphics[scale=0.4]{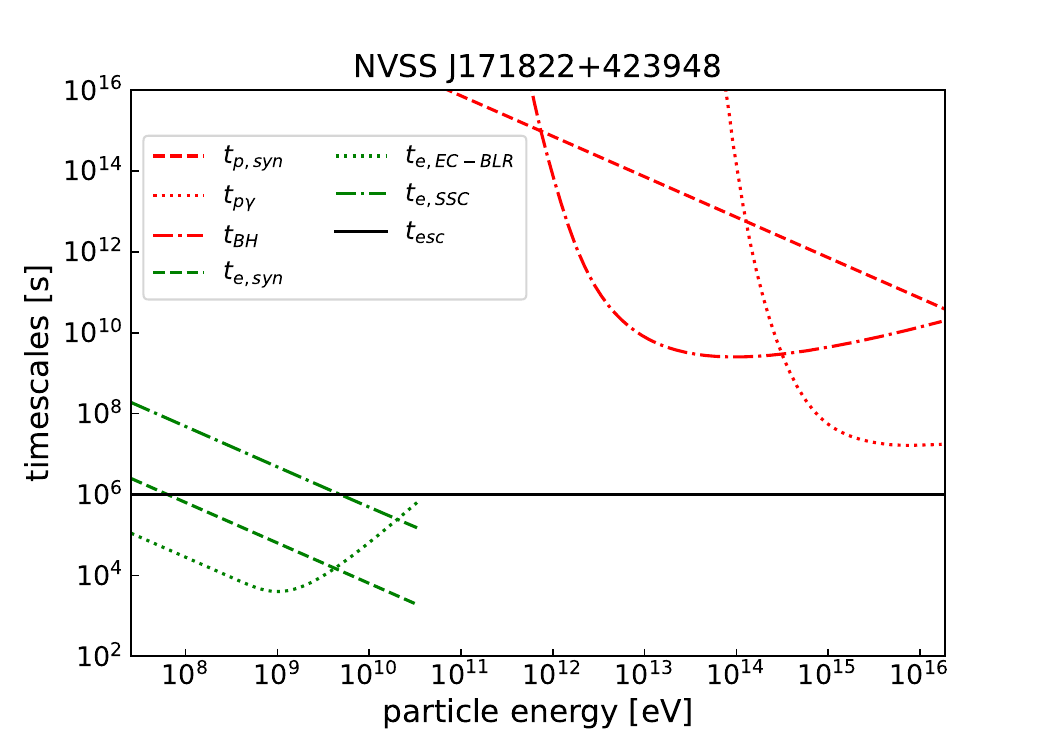}
    \caption{ The time scales for different cooling processes of electrons (green curve) and protons (red curve) within the blob are presented as a function of particle energy. The relevant parameters are taken from Table \ref{t1}. Both the particle energy and time scales are evaluated in the jet's comoving frame. The black horizontal line indicates the particles' escape time scale.
}
    \label{f1}
\end{figure*}

5. The value of $\gamma_{p,max}$ can be determined by considering that the proton acceleration time is equal to  the minimum of the energy loss and escape timescales. Assuming that the protons acceleration takes place at the diffusive shocks, the expression for the acceleration timescale can be written as
\begin{equation}
t_{acc} =  \frac{\eta\gamma_pm_pc}{eB},
\end{equation}
where $\eta \geq 1$  represents the acceleration efficiency, with $\eta =1$ corresponding to the maximum possible acceleration rate, which occurs when diffusion is in the Bohm limit \citep{2019MNRAS.489.4347O}. The maximum proton energy is determined by the fastest cooling mechanism and is obtained by  $t_{acc}(\gamma_{p,max})$ = $\min[t_{p,syn}(\gamma_{p,max}), t_{p\gamma}(\gamma_{p,max}), t_{BH}(\gamma_{p,max}), t_{dyn}]$. In Figure \ref{f1}, it can be seen that $\min[t_{p,syn}(\gamma_{p,max}), t_{p\gamma}(\gamma_{p,max}), t_{BH}(\gamma_{p,max}), t_{dyn}]$ = $t_{dyn}$. Then, the maximum proton Lorentz factor is given by
\begin{equation}
\gamma_{p,max} = \frac{eBR}{\eta{m}_pc^2} = 3.2 \times {10}^9 {(\eta/1)}^{-1} (\frac{B}{1 G}) (\frac{R}{{10}^{16}cm}),
\end{equation}
In our model, if we consider $\eta = 1$, the maximum proton energy reaches $E_p = \gamma_{p,max}m_p c^2 \approx 10$ EeV. However, a maximum proton energy higher than a few EeV would produce a neutrino spectrum that peaks beyond 100 PeV, and the sub-PeV neutrino flux would be highly suppressed \citep[e.g.,][]{2014PhRvD..90b3007M,2018ApJ...864...84K, 2019MNRAS.489.4347O}.
Additionally, in modeling the neutrino emission from FSRQs, \cite{2019ApJ...871...41P} suggested that a low acceleration efficiency of $\eta = 10^3$  would result in neutrinos with a maximum energy of $\sim$ 10 PeV in the AGN frame, which aligns with the maximum energy of the observed astrophysical neutrinos. Consequently, we consider $\eta = 10^3$, where protons cannot be effectively accelerated.

6. We use the  observed minimum variability time-scale $t^{obs}$ to  constrain the radius of the
emission region as $R \leq (\delta c t^{obs})/(1+z)$. For  NVSS J171822+423948, the ZTF $i$-band flux shows variability within 10 days, while for GB6 J2113+1121, the dense monitoring of WISE at MJD 59,146 has showed rapid variability within $\sim$ 12 hr.

7. We consider a conical jet with an opening angle of $ \sim 1/\delta$, and if the radiation region fully occupies the entire open jet, the distance from the radiation region to the central black hole can be estimated to be $r_{in}^{'} \approx R/(1/\delta)$.

Finally, we are left with nine free parameters: $R$, $B$, $\delta$, $L_{e,inj}$, $\gamma_{e,break}$, $\gamma_{e,max}$, $n_{e,1}$, $n_{e,2}$, and $L_{p,inj}$. We search for parameters that adequately describe the multi-wavelength data. Our goal is not to obtain the best model parameters by fitting the data through minimizing the likelihood function, but rather to maximize the neutrino flux while explaining the electromagnetic spectrum (SED). Therefore, we choose to "fit by eye," manually adjusting certain parameters to match the modeling results with the electromagnetic SED, while ensuring a reasonable model, and obtain the neutrino flux. It should be noted that the $\gamma-$ray data for GB6 J2113+1121 and NVSS J171822+423948 are taken from the averages during their respective $\gamma-$ray flaring periods. However, due to observational limitations, the optical, infrared, and X-ray data are not average values that match the $\gamma-$ray flaring periods, but rather values from a specific observation or period within the $\gamma-$ray flaring periods. This means that the data collected for the two sources were not obtained simultaneously, which weakens the use of a steady-state solution.

\section{Modeling results} \label{sec:3}

In this  section, we present the results of  the
one-zone leptohadronic model applied to GB6 J2113+1121 and  NVSS J171822+423948, including cooling timescales related to electrons and protons, the SEDs of multi-band electromagnetic and neutrino emissions, neutrino detection rates, baryon loading factor, and neutrino to-$\gamma$-ray luminosity ratio.

The timescales of various cooling processes for electrons and protons are shown in Figure \ref{f1}. It can be observed that for the primary electrons, the SSC cooling time is relatively longer than the EC cooling time, indicating that high-energy $\gamma-$rays are primarily produced by EC emission. This is because we assume that the radiation regions of both GB6 J2113+1121 and NVSS J171822+423948 are located near the BLR boundary (i.e., $r_{in}^{'} \approx r_{BLR}^{'}$), where the boosted external photon field leads to rapid cooling of the primary electrons. The proton escape timescale $t_{dyn}$ is shorter than the proton synchrotron radiation timescale $t_{p,syn}$, the $p\gamma$ energy loss timescale $t_{p\gamma}$, and the BH pair-production cooling timescale $t_{BH}$, which means that protons cannot effectively cool. At low energies, protons primarily lose energy through BH pair production. As the proton energy increases, the $p\gamma$ process becomes increasingly significant, and protons mainly lose energy through the $p\gamma$ process.

\begin{figure*}
    \centering
    \includegraphics[scale=0.4]{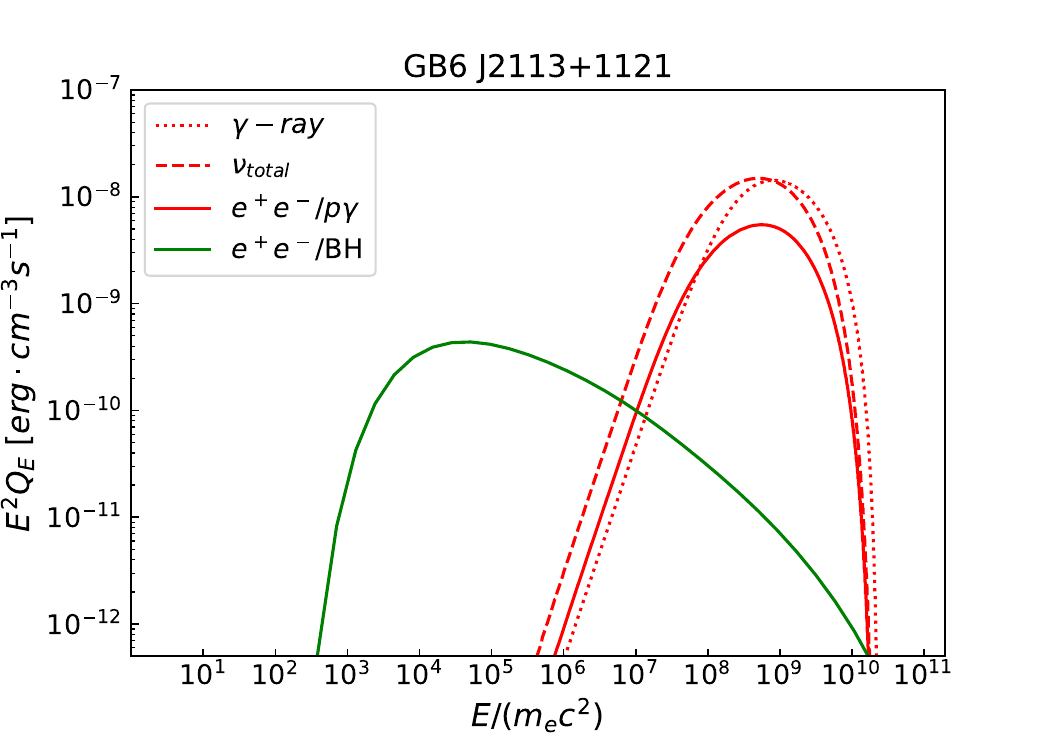}
    \includegraphics[scale=0.4]{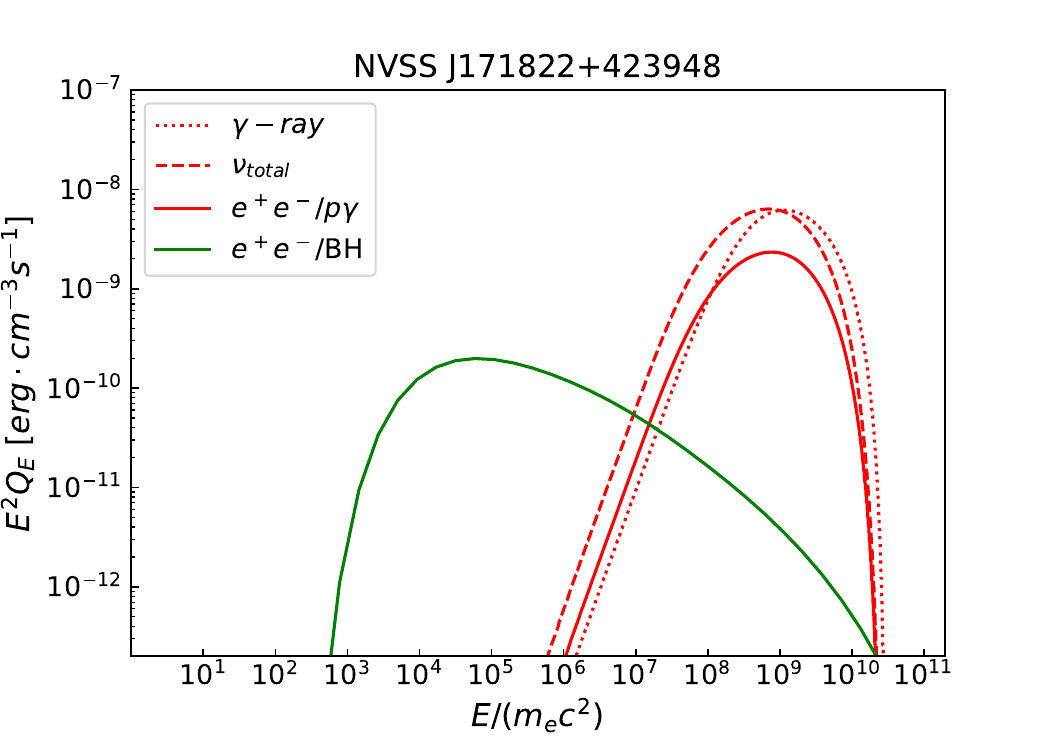}
    \caption{The injection spectra of secondary particles produced by hadronic processes. The solid green line represents the pair spectra from the Bethe-Heitler process, the dotted red line denotes the $\gamma$-ray spectra from neutral pion decay, the solid red line indicates the pair spectra from charged pion decays, and the dashed red line represents the all-flavor neutrino spectra from charged pion decays. All quantities are measured in the comoving frame of the jet.}
    \label{f2}
\end{figure*}

\begin{figure*}
    \centering
    \includegraphics[scale=0.4]{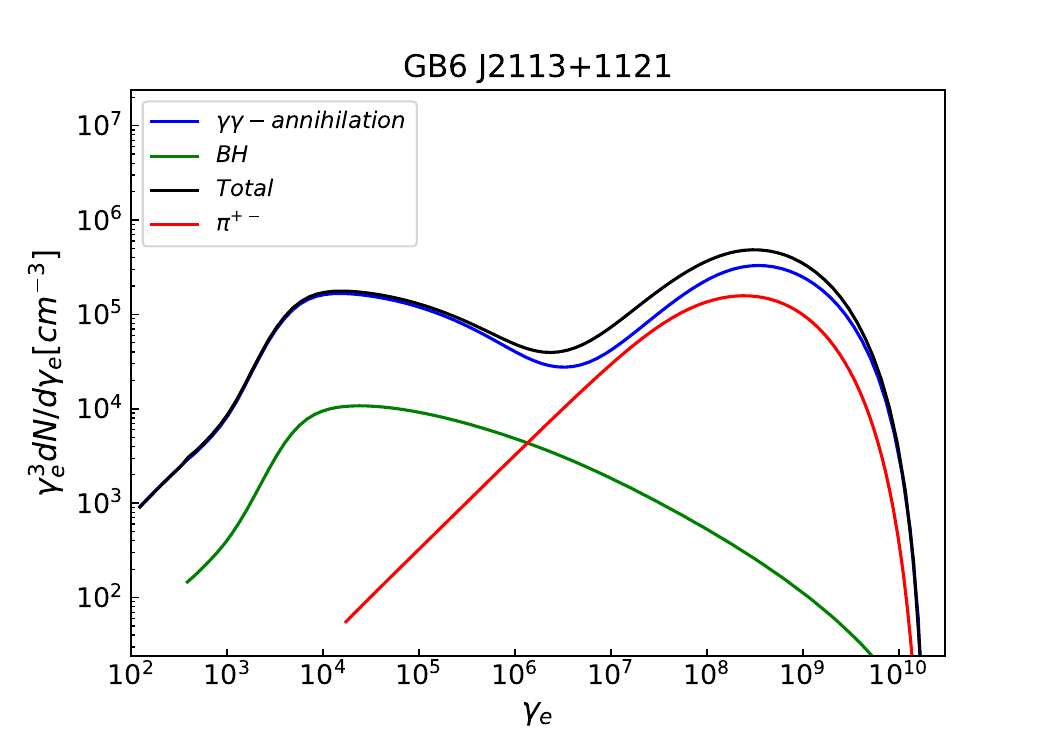}
    \includegraphics[scale=0.4]{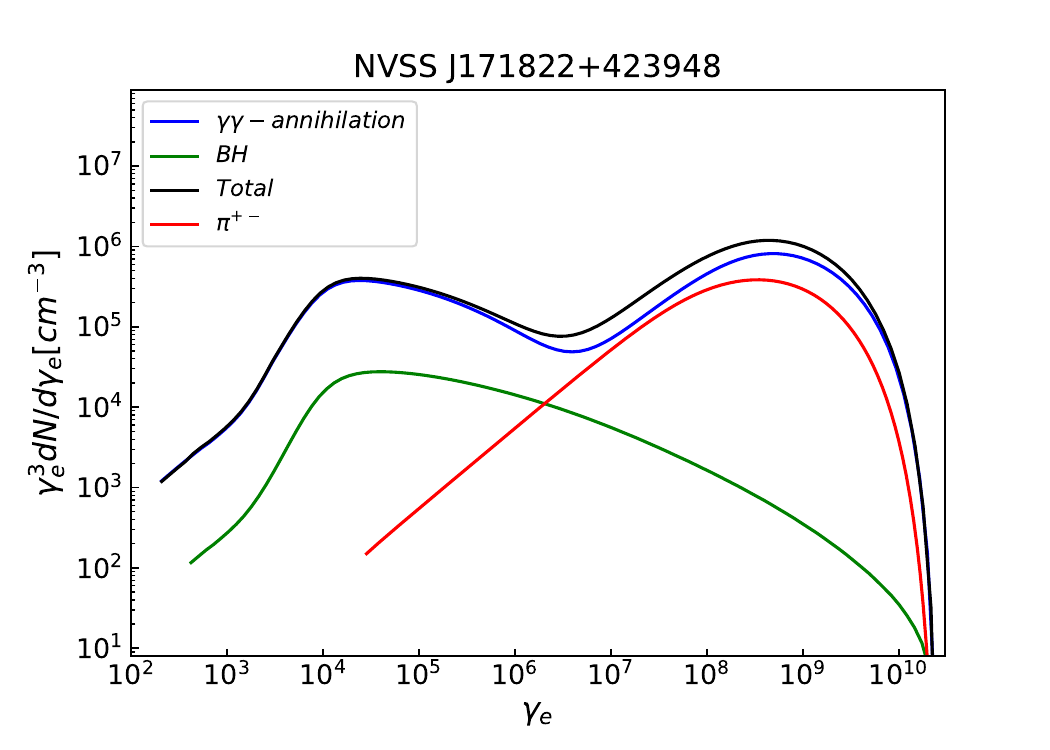}
    \caption{The distribution of emitting electron-positron pairs produced by hadronic processes in GB6 J2113+1121 and NVSS J171822+423948. 
The green curve represents the emitting pairs distribution from the Bethe-Heitler process, the red curve represents the emitting pairs distribution obtained from the $\pi^{\pm}$ decay. The blue curve represents  the emitting pairs distribution from $\gamma\gamma$ annihilation, primarily originating from the $p\gamma$ component, while the contribution from the Bethe-Heitler process can be neglected. And the black curve represents the total emitting positron-electron pairs distribution produced by hadronic processes. All quantities are measured in the comoving frame of the jet.}
    \label{f3}
\end{figure*}

It can be seen that the first generation of lower-energy pairs are primarily produced by the Bethe-Heitler process, while pairs from the $p\gamma$ process have much higher energies (see Figure \ref{f2}). However, the distribution of emitting electron-positron pairs produced by hadronic processes, as shown in Figure \ref{f3} , indicates that for both GB6 J2113+1121 and NVSS J171822+423948, the emitting electron-positron pairs are primarily contributed by the $p\gamma$ process, even at lower energies. This is due to the dense soft photon field, which leads to significant absorption of high-energy $\gamma-$ rays produced by the $p\gamma$ process, resulting in electron-positron pairs that overshadow those from the Bethe-Heitler process.

\begin{figure*}
    \centering
    \includegraphics[scale=0.7]{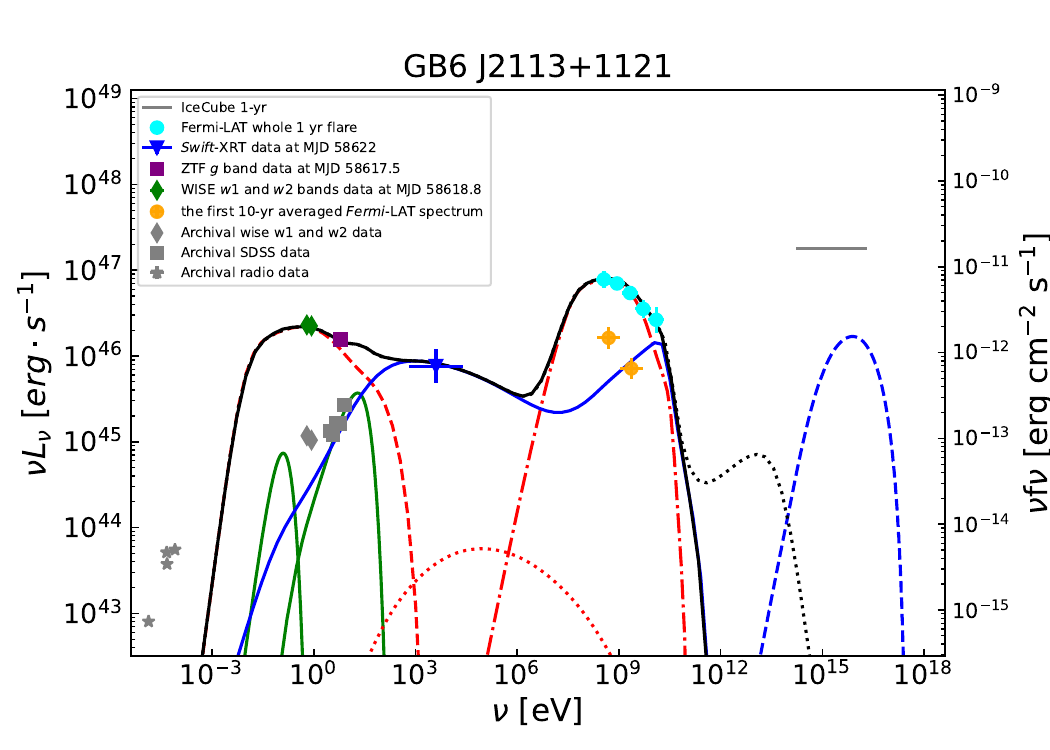}
    \includegraphics[scale=0.7]{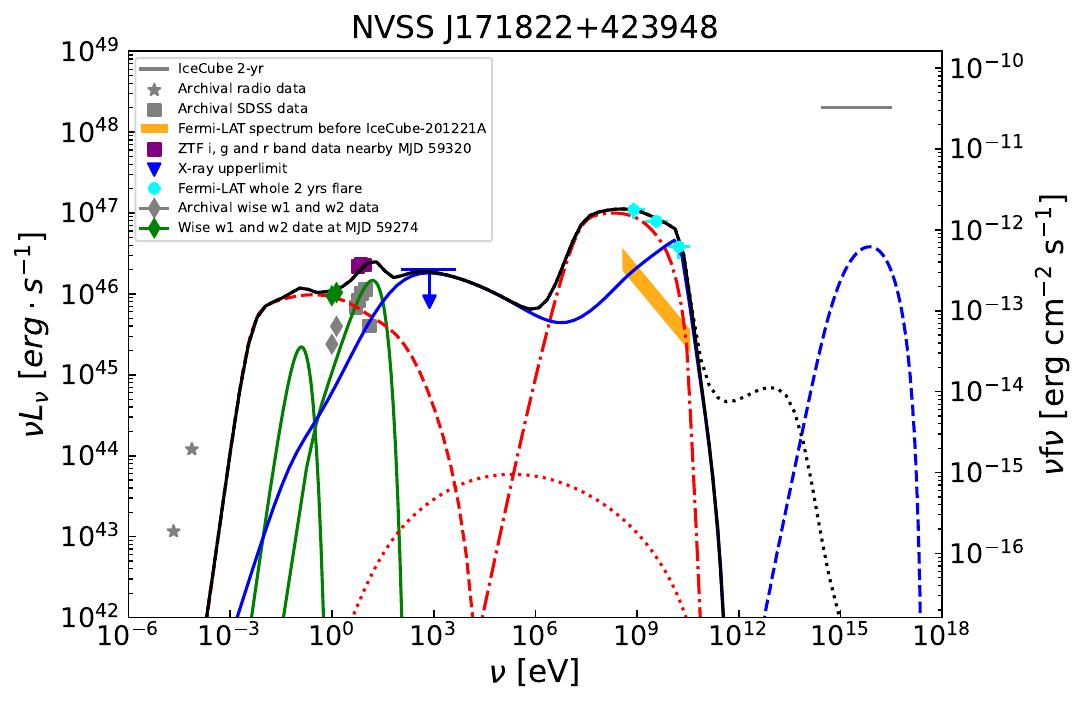}
    \includegraphics[scale=0.8]{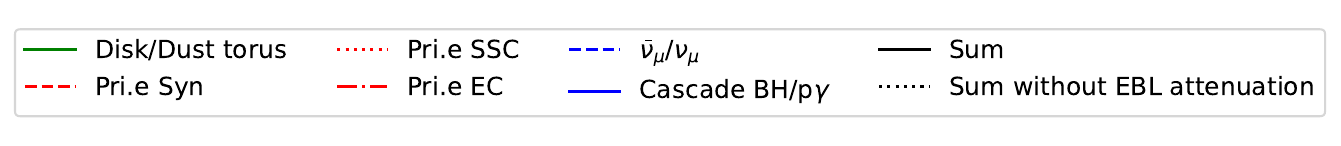}
    \caption{The broadband SED of GB6 J2113+1121 and NVSS J171822+423948 during their respective $\gamma$-ray high-state periods. For GB6 J2113+1121, all data points are from \citep{2022ApJ...932L..25L}, while for NVSS J171822+423948, all data points are from \citep{2024ApJ...965L...2J}. Here, we do not consider the radiation from the dust torus as a target photon field. The red dashed curve represents the synchrotron emission from primary electrons, the red dotted curve represents the SSC emission from primary electrons, and the red dotted-dashed curve represents the EC emission from primary electrons. The solid blue curve represents the emission from pair cascades produced by hadronic processes. The green curve indicates the emission from the accretion disk and the dusty torus. The solid black curve is the total electromagnetic emission from the blob, while the dotted black curve is the total electromagnetic emission from the blob without EBL attenuation. The blue dashed curve shows the $\bar{\nu_\mu}/\nu_\mu$ neutrino energy spectrum. The horizontal gray line represents the differential muon neutrino flux required to produce one GFU alert event at the declination of GB6 J2113+1121/NVSS J171822+423948 during their respective $\gamma$-ray high state periods, assuming an $E^{-2}$ neutrino spectrum. All of these quantities are measured in the cosmological rest frame.}
    \label{f4}
\end{figure*}

\begin{figure*}
    \centering
    \includegraphics[scale=0.8]{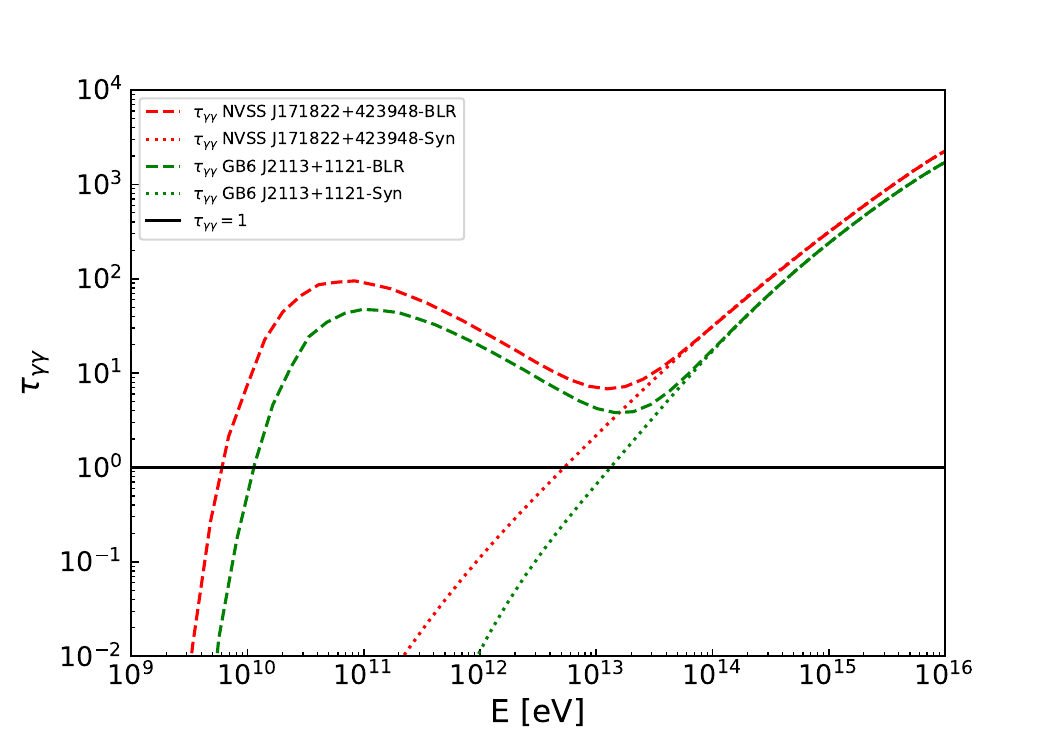}
    \caption{The optical depth of $\gamma\gamma$ annihilation for GB6 J2113+1121 and NVSS J171822+423948, with the x-axis representing the $\gamma$-ray energy in the observer's frame. The dashed line represents the optical depth when considering the soft photon field from the external photon field of BLR and the synchrotron radiation of primary electrons, while the dotted line indicates the optical depth when only considering the soft photon field from the synchrotron radiation of primary electrons. The black horizontal line indicates an optical depth equal to 1.}
    \label{f5}
\end{figure*}

The SED resulting from our one-zone lepto-hadronic modeling of the two sources, including both the photon and neutrino components, is shown  in Figure \ref{f4}. Both GB6 J2113+1121 and  NVSS J171822+423948 are classified as LSPs,  with their synchrotron emission peaks below $1 \times 10^{14} $ Hz. In GB6 J2113+1121, the IR-optical emissions result from the synchrotron radiation of primary electrons. In  NVSS J171822+423948 , the infrared emission also originates from the synchrotron radiation of primary electrons. However, due to its bright accretion disk luminosity, the optical emission of  NVSS J171822+423948 is a combination of synchrotron radiation from primary electrons and thermal radiation from the accretion disk. At higher energies, the X-ray emissions from both sources are dominated by synchrotron radiation from pair cascades, with emissions peaking around $\sim$ 1 keV for  GB6 J2113+1121 and $\sim$ 0.5 keV for NVSS J171822+423948 in the observer’s frame. This result is similar to the modeling results of \cite{2024A&A...681A.119R}, which indicate that the X-rays of some FSRQs can be dominated by pair cascades. This differs from leptonic models, where hard X-rays are typically produced by SSC radiation of the primary electrons. The X-ray flux at keV energies strictly constrains the pair cascades emission,  thereby limiting the neutrino flux (see also \cite{2023ApJ...958L...2F}). Due to the lack of available X-ray data for NVSS J171822+423948, we consider its X-ray luminosity to be $\sim$ 2 × 10$^{46}$ erg s$^{-1}$, a typical value for the high-redshift $\gamma-$ray FSRQs\citep{2017ApJ...837L...5A}.
The hundreds of MeV to $\sim$ 5 GeV $\gamma$ rays from these two sources primarily originate from the EC emission of the primary electrons. However, we found that the synchrotron emission of pair cascades produced by $p\gamma$ process from both sources can significantly contribute to the radiation above $\sim$ 5 GeV, which, like X-rays, in turn limits the neutrino flux. For both GB6 J2113+1121 and NVSS J171822+423948, the spectrum of the high-energy part of the steady-state secondary positron-electron pairs peaks at $\gamma_e \sim 1 \times 10^9$ (see Figure \ref{f3}), and their synchrotron peak energy is given by $E_{s}^{obs} = (B/B_{cr})m_ec^2\gamma_e^2\delta/(1+z) \gtrsim$ 100 GeV, where $B_{cr} = 4.414 \times 10^{13}$ G is the critical magnetic field \citep{2008ApJ...686..181F} and $z$ is the redshift of the sources.   However, the contribution of cascade emission to $\gamma$ rays is mainly in a few GeV range, rather than at higher energies. This is because the intense BLR photon field leads to significant absorption of high-energy $\gamma$ rays produced by the cascade process that exceed a few tens of GeV. Figure \ref{f5} shows the $\gamma$-ray optical depth within the two sources both with and without the presence of the  BLR photon field. In the presence of a BLR photon field, for GB6 J2113+1121 and  NVSS J171822+423948, an optical depth of $\tau_{\gamma\gamma} = 1$ corresponds to $\gamma$-ray energies of $\sim$ 10 GeV and $\sim$ 5 GeV in the observer’s frame, respectively. The significant absorption of $\gamma$-rays by the BLR weakens the connection between neutrinos and the $\gamma$-rays produced by electromagnetic cascades \citep{2021ApJ...911L..18K}. 
Similarly, a spectral break in the $\gamma$-ray spectrum at a few GeV may indicate an effective environment for neutrino production.
Additionally, considering that electromagnetic cascades inevitably boost X-rays, it suggests that the correlation between X-rays and neutrinos may be stronger than that between $\gamma$-rays and neutrinos.

Table \ref{t1}  shows the results of the parameters obtained from modeling the electromagnetic emission and neutrino emission. The magnetic field strengths for GB6 J2113+1121 and NVSS J171822+423948 are 6 G and 2.5 G, respectively, both of which are approximately one order of magnitude larger than that of TXS 0506+056 \citep{2018ApJ...864...84K}. SED modeling studies suggest that a magnetic field strength of a few Gauss is characteristic of FSRQs \citep[e.g.,][]{2010MNRAS.402..497G,2024A&A...681A.119R}. This can be attributed to the Doppler-enhanced external photon field from the BLR, which requires a relatively strong magnetic field to enable efficient synchrotron emission of electrons, thereby explaining the low-energy hump in the SED.  We found that for both GB6 J2113+1121 and NVSS J171822+423948, the proton injection luminosity is two orders of magnitude higher than the electron injection luminosity ($L_{p,inj}/L_{e,inj} \geq 100$), suggesting that a proton-dominated jet scenario may be necessary to explain the IceCube alert neutrino fluxes. The energies of the neutrino events IceCube-191001A and IceCube-201221A are both $\sim$ 0.2 PeV. Through the relationship between neutrino energy and proton energy, the parent proton energy in the jet's comoving frame can be expressed as $E_p=20E_{\nu}^{obs}(1+z)/\delta$. For GB6 J2113+1121 and  NVSS J171822+423948, the proton energies are $\sim$ 0.3 PeV and $\sim$ 0.6 PeV, respectively, with corresponding proton Lorentz factors of $\sim$ 3 $\times$ 10$^5$ and $\sim$ 6 $\times$ 10$^5$. Although our model uses maximum proton Lorentz factors that are much higher than these values, with  $\gamma_{p,max}$ = 2 $\times$ 10$^7$ for  GB6 J2113+1121 and $\gamma_{p,max}$ = 2.5 $\times$ 10$^7$ for  NVSS J171822+423948, the peak neutrino flux for  NVSS J171822+423948 remains around $E_{\nu}^{obs} \sim$2 PeV and for GB6 J2113+1121 remains around $E_{\nu}^{obs} \sim$3 PeV  in the observer's frame, both of which still fall within the range of 100 TeV to 10 PeV, roughly corresponding to IceCube's detection sensitivity range \citep{2016ApJ...833....3A}.  For GB6 J2113+1121, the peak $\nu_\mu/\bar{\nu_\mu}$ flux is $\sim$ $2 \times 10^{-12}$ erg cm$^{-2}$ s$^{-1}$, while for NVSS J171822+423948, the peak $\nu_\mu/\bar{\nu_\mu}$ flux is $\sim$ $8 \times 10^{-13}$ erg cm$^{-2}$ s$^{-1}$ (see Figure \ref{f4}).
 The number of muon neutrinos detected by IceCube over time $\triangle T^{obs}$ is:
\begin{equation}
     {N_{\nu_{\mu}^{obs}}}  = \int_{\epsilon_{\nu_{\mu, \text{min}}^{obs}}}^{\epsilon_{\nu_{\mu, \text{max}}^{obs}}} {A_{\text{eff}}(\epsilon_{\nu_{\mu}^{obs}}, \delta) \frac{dN_{\nu_\mu^{obs}}}{d\epsilon_{\nu_\mu^{obs}}} \Delta{T}^{obs}} \, d\epsilon_{\nu_{\mu}^{obs}},
\end{equation}
where ${\epsilon_{\nu_{\mu, min}^{obs}}} = 80$ TeV and ${\epsilon_{\nu_{\mu, max}^{obs}}} = 8$ PeV represent the range of energies within which 90$\%$ of neutrinos are expected to be found in the Gamma-ray Follow-Up (GFU) channel \citep{2023ApJS..269...25A}.  $A_{\text{eff}}(\epsilon_{\nu_{\mu}^{obs}}, \delta)$ is the energy- and declination-dependent effective area of IceCube, and $\frac{dN_{\nu_\mu^{obs}}}{d\epsilon_{\nu_\mu^{obs}}} = \frac{1}{3} \frac{dN_{\nu^{obs}}}{d\epsilon_{\nu^{obs}}}$ represents the differential flux of muon neutrinos. Here, $\frac{dN_{\nu^{obs}}}{d\epsilon_{\nu^{obs}}}$ is the all-flavor neutrino and anti-neutrino flux (differential in energy) obtained from the modeling results, with the factor of 1/3 accounting for neutrino oscillations in vacuum. The time period $\triangle T^{obs}$ is considered to be equal to the $\gamma-$ray flare duration of each source.
For GB6 J2113+1121 , the expected number of $\bar{\nu_\mu}/\nu_\mu$ neutrinos during the $\sim$ 1 years of $\gamma$-ray flaring activity is about 0.02. For  NVSS J171822+423948, the expected number of $\bar{\nu_\mu}/\nu_\mu$ neutrinos during the $\sim$ 2 years of $\gamma$-ray flaring activity is $\sim$ 0.008, which is consistent with recent findings on the neutrino candidate 3HSP J095507.9 +355101 \citep{2020ApJ...899..113P}. The total number of neutrinos produced is treated as the mean number of events in a Poisson distribution, and the probability of detecting neutrinos from the $\gamma-$ray flare is \citep{2024ApJ...977...42R}:
\begin{equation}
P_{Det} = P(k\geq1) = 1-P(0) = 1 - e^{-N_{\nu_\mu^{obs}}}
,\end{equation}
where $k$ is the number of neutrinos detected by IceCube. The Poisson probability of detecting one neutrino is $\sim$ 2 $\%$ for GB6 J2113+1121 and $\sim$ 0.8 $\%$ for NVSS J171822+423948, which could be interpreted as a statistical fluctuation that accounts for the association.
The expected number of neutrinos from both sources is very low, primarily because they are both high-redshift sources and thus very distant from Earth. In particular,  NVSS J171822+423948 has a redshift as high as 2.7. 
In addition, for NVSS J171822+423948, the effective area of IceCube is relatively low, measuring only about 6 m$^2$ for the neutrino IceCube-201221A with an energy of 0.174 TeV.
In our calculations, we consider the neutrino emission duration for both sources to be equal to the $\gamma$-ray flaring duration. However, this duration is much longer than the time it takes for the blob to cross the BLR, which is $\sim$ $r_{in}^{'}/(\delta^2 c)(1+z)$ in the observer's frame.  If the neutrino emission duration were to match the time it takes for the blob to traverse the BLR, the neutrino detection probability for both sources would be very low. However, if multiple blobs, similar to the one described in our model, are continuously formed in or near the BLR over a period of several years,  assuming the radiation originates from a region located at a fixed distance $r_{in}^{'}$ from the central black hole, which remains constant throughout the modeled period, i.e., the emission region is considered a stationary feature in the jet, this could lead to neutrino emission lasting on the order of years\citep{2019ApJ...886...23X}.

Then, we also calculated the total jet power for both sources separately. The kinetic  luminosity of relativistic particles  can be described as follows.
\begin{equation}
L_{i,k}^{'} = \pi R^2\Gamma^2cm_ic^2\int N_i(\gamma_i)\gamma_i \, d\gamma_i
,\end{equation}
where $i = e, p$.
Consider a two-sided jet, the total jet kinetic luminosity is estimated as $L_{p,jet}^{'}$ = 2 $\times$ ($L_{e,k}^{'}$ + $L_{p,k}^{'}$ + $L_{B,jet}^{'}$), where $L_{B,jet}^{'} = \pi R^2\Gamma^2c {B^2}/(8\pi)$ is the magnetic luminosity. The Eddington luminosity is given by  $L_{Edd}^{'}$ = 1.26 $\times$ 10$^{46}$ ${M^{'}_{BH}}/({10^8M_{\odot}})$ erg s$^{-1}$, where $M_{\odot}$ is the solar mass. For GB6 J2113+1121, the required jet luminosity is about 2.2 $\times 10^{47}$ erg s$^{-1}$, which is $\sim$ 10 times greater than  the Eddington luminosity of this source $L_{Edd}^{'} =$ 2$\times 10^{46}$ erg s$^{-1}$, considering the black hole mass of this source $M_{BH}^{'}$ = $1.6 \times 10^8 M_{\odot}$ \citep{2022ApJ...932L..25L}. Similarly, for NVSS J171822+423948, the required jet luminosity is about $5.5 \times 10^{47}$ erg s$^{-1}$, which is $\sim$ 4 times greater than the Eddington luminosity of this source $L_{Edd}^{'} =$ $1.3 \times 10^{47}$ erg s$^{-1}$, considering the black hole mass of this source $M_{BH}^{'}$ = $10^9 M_{\odot}$ \citep{2024ApJ...965L...2J}. For both GB6 J2113+1121 and NVSS J171822+423948, the required jet luminosities exceed their respective Eddington luminosities, which is consistent with earlier studies on blazars \citep{2014Natur.515..376G}. Additionally, considering that flares are just a short period of activity in the life of AGNs, the case of jet luminosities temporarily exceeding Eddington luminosities is acceptable.

Based on the modeling results of the SED, we can determine the baryon loading factor, which is defined as $\xi = L_{p}^{obs} / L_{\gamma}^{obs}$. Here, $L_{\gamma}^{obs}$ represents the observed $\gamma$-ray luminosity integrated over the 100 MeV-300 GeV energy range, roughly corresponding to the Fermi-LAT
observation range, while $L_{p}^{obs} =\delta^4L_p \approx (4/3)\Gamma^2 L_{p,k}^{'} $ denotes the isotropic-equivalent proton luminosity in the observer frame. For GB6 J2113+1121,   $L_{\gamma}^{obs} \approx 3.0\times10^{47}$ erg s$^{-1}$ and $L_{p}^{obs} \approx 1.2\times10^{50}$ erg s$^{-1}$, resulting in a baryon loading factor  $\xi \approx 4.0 \times 10^2$. For  NVSS J171822+423948, $L_{\gamma}^{obs} \approx 4.8 \times10^{47}$ erg s$^{-1}$ and $L_{p}^{obs} \approx 1.4 \times10^{50}$ erg s$^{-1}$, yielding a baryon loading factor $\xi \approx 2.9 \times 10^2$. The baryon loading factors for both sources are consistent with the results proposed by \cite{2019ApJ...871...41P} for FSRQs with $L_{\gamma}^{obs} \approx 5.0\times10^{47}$ erg s$^{-1}$, where they predicted a baryonic loading of about a few hundred.  In contrast, the baryon loading factors for both sources are much lower than those obtained from the SED modeling of BL Lacs \citep{2015MNRAS.448.2412P}, where only the jet synchrotron photons are considered as targets for photohadronic interactions. However, the baryon loading factors for both sources are comparable to the value estimated for TXS 0506+056 during its 2017 $\gamma$-ray flare \citep{2018ApJ...864...84K}. The all-flavor neutrino to-$\gamma$-ray luminosity ratio $Y_{\nu\gamma} = L_{\nu+\bar{\nu}}^{obs}/L_{\gamma}^{obs}$ is $\sim$ 0.5 for GB6 J2113+1121 and $\sim$ 0.7 for  NVSS J171822+423948, which is about one order of magnitude higher than the value $Y_{\nu\gamma}= 0.03$ observed for TXS 0506+056 during the 2017 flare \citep{2018ApJ...864...84K}. Considering that $Y_{\nu\gamma} \sim (3/8)f_{p\gamma}\xi$, where $f_{p\gamma}$ is the $p\gamma$ production efficiency, this implies that the neutrino production efficiency of  GB6 J2113+1121 and NVSS J171822+423948 is higher than that of TXS 0506+056. This is because, although TXS 0506+056 also introduces an external photon field, its energy density is very low, with $u_{ext} = 0.033 $ erg cm$^{-3}$ \citep{2018ApJ...864...84K}. Moreover, the baryon loading factors of GB6 J2113+1121 and NVSS J171822+423948 are lower than that of the X-ray flare-selected neutrino candidate 3HSP J095507.9+355101, while their $Y_{\nu\gamma}$ values are similar to that of 3HSP J095507.9+355101 (see Models A-D, \citep{2020ApJ...899..113P}). This is because the modeling of 3HSP J095507.9+355101 does not account for an external photon field, requiring a higher proton luminosity to produce neutrinos. Meanwhile, the hadronic cascade component of 3HSP J095507.9+355101 makes a non-negligible contribution to its high-energy emission. Compared to another neutrino candidate, PKS 0735+178, our baryon loading factors are close to the 
$\xi \approx 9.0 \times 10^2$ obtained from the hybrid-ext model of PKS 0735+178, which incorporates an external photon field from the BLR \citep{2023MNRAS.519.1396S}. However, the ratio of all-flavor neutrino luminosity to $\gamma-$ray luminosity for PKS 0735+178 is $\sim$ $6.9 \times 10^{-3}$, significantly lower than that of GB6 J2113+1121 and NVSS J171822+423948. This discrepancy arises because the hadronic cascade component's contribution to high-energy emission in PKS 0735+178 is negligible. A similar situation is seen in the SED modeling of the neutrino candidate PKS B1424-418, where the ratio of all-flavor neutrino luminosity to $\gamma-$ray luminosity is low, on the order of 5$\%$ \citep{2017ApJ...843..109G}.

\section{Discussion} \label{sec:4}

Since both GB6 J2113+1121 and NVSS J171822+423948 are FSRQs with very high accretion disk luminosities and both exhibit short-term variability, which implies a compact emission region, in our model, we assume that the radiation regions of both sources are located close to the BLR boundary, i.e., $r_{in}^{'} \approx r_{BLR}^{'}$. Therefore, when calculating the neutrino production for these two sources, we include the external photon field from the BLR as the target photon field for the $p\gamma$ process. In contrast, for another well-known neutrino candidate with a similarly high accretion disk luminosity, PKS B1424-418, \cite{2017ApJ...843..109G} in their multimessenger modeling of this source consider the neutrino emission region to be located outside the BLR, and the photon field for the $p\gamma$ interaction comes solely from the synchrotron radiation of the primary electrons. Moreover, some studies indicate that certain FSRQs have a compact emission region but also exhibit radiation well beyond the BLR, with a notable example being the FSRQ PKS 1222+216 \citep{2011A&A...534A..86T}. This suggests that short-term variability alone is not sufficient to conclusively prove that the emission region is located within the BLR. Therefore, it is necessary to study the impact of different emission region locations on the neutrino production efficiency in GB6 J2113+1121 and NVSS J171822+423948. Here, we consider three scenarios: the emission region is located within the BLR, beyond the BLR but still inside the dusty torus, and outside the dusty torus, while keeping the radius of the emission region and the Doppler factor the same as in the BLR scenario. For the latter two cases, we do not perform additional multi-messenger SED modeling but instead study the neutrino production efficiency using the reaction efficiency of the $p\gamma$ process, $f_{p\gamma}$. We propose that the electromagnetic radiation from GB6 J2113+1121 and  NVSS J171822+423948 can be explained using a different set of electron and magnetic field parameters (with the radius of the radiation blob and Doppler factor unchanged). Therefore, we use the synchrotron radiation from primary electrons within the BLR to represent the synchrotron radiation from primary electrons in the cases where the radiation region is located inside and outside the dust torus. 

\begin{figure*}
    \centering
    \includegraphics[scale=0.5]{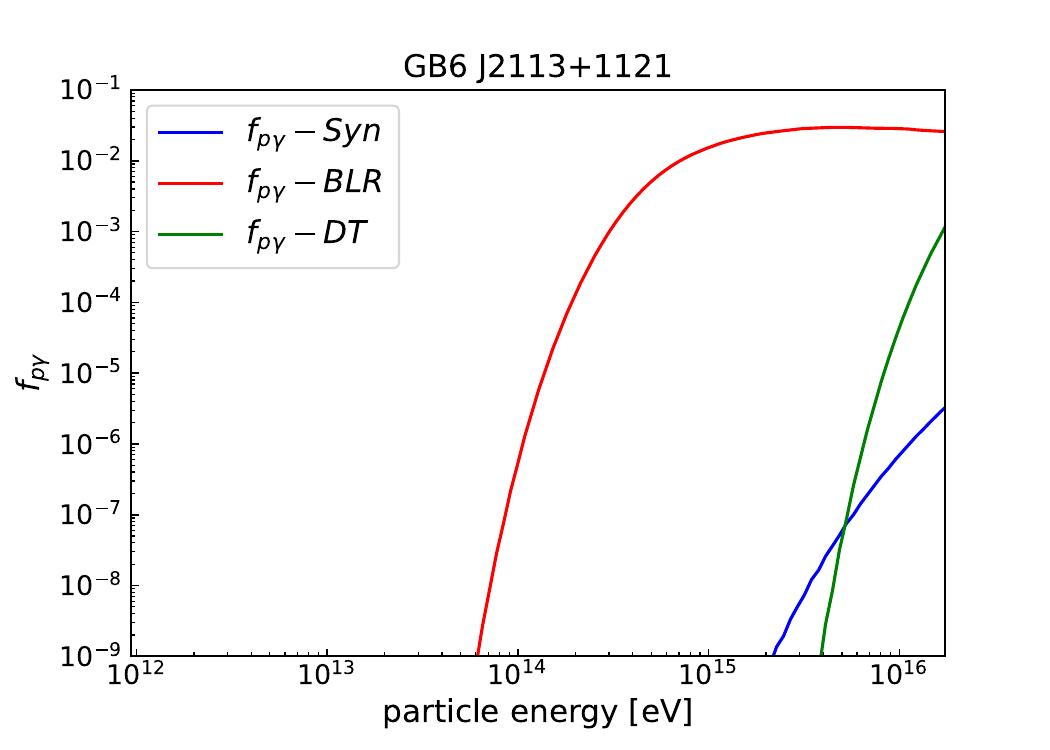}
    \includegraphics[scale=0.5]{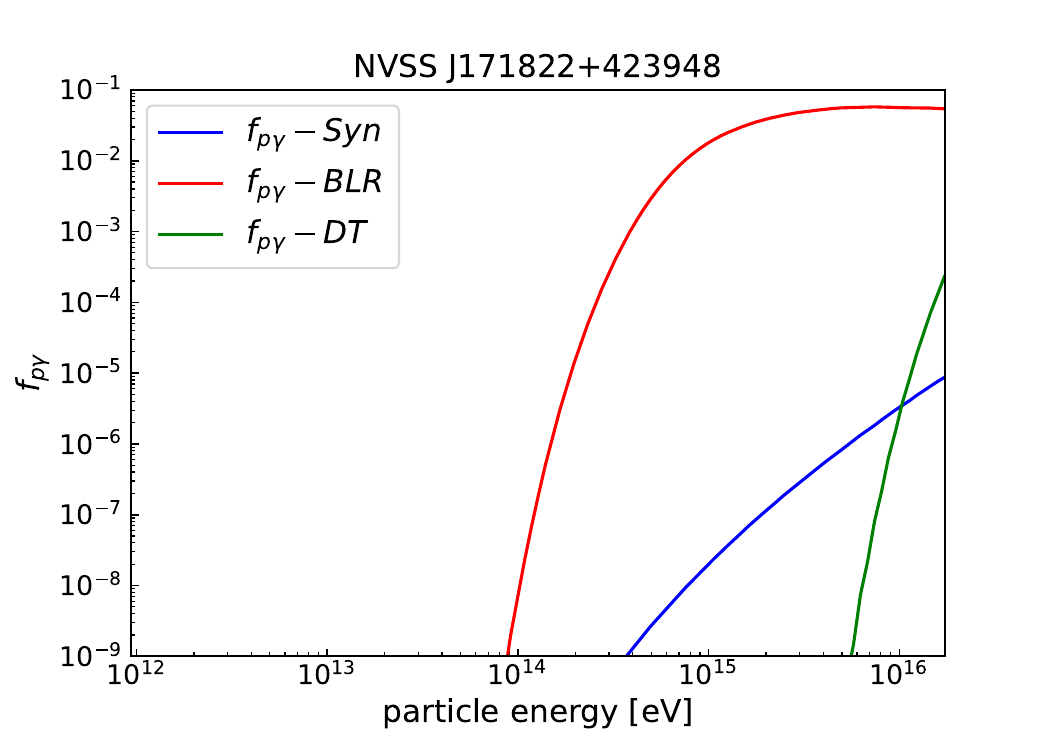}
    \caption{The efficiency of $p\gamma$ reactions in different soft photon fields is shown. The red curve represents the $p\gamma$ reaction efficiency related to proton energy when the radiation from the BLR acts as the external photon field. The blue curve represents the $p\gamma$ reaction efficiency when the photon field from synchrotron radiation of the primary electrons in the jet serves as the soft photon field. The green curve indicates the $p\gamma$ reaction efficiency when the radiation from the DT acts as the soft photon field. All quantities are measured in the comoving frame of the jet.}
    \label{f6}
\end{figure*}

The radiation from the dust torus in the jet comoving frame is modeled as described in section \ref{subsec:2-1}. The proton energy-loss timescale is given by
\begin{equation}
t_{cool}^{-1} \equiv t_{dyn}^{-1} + t_{p,syn}^{-1} + t_{p\gamma}^{-1} + t_{BH}^{-1}
,\end{equation}
The fraction of energy converted into pions is given by $f_{p\gamma} \equiv t_{cool}/t_{p\gamma}$. However, from Figure \ref{f1}, we can see that $t_{dyn}$ is  smaller than the other energy loss timescales($t_{p,syn}$, $t_{p\gamma}$, $t_{BH}$), so we have $f_{p\gamma} \approx t_{dyn}/t_{p\gamma} $. According to \cite{2021JCAP...10..082O}, the all-flavour neutrino luminosity  in the jet-comoving frame is given by 
\begin{equation}
\epsilon_{\nu}L_{\epsilon_{\nu}} \approx \frac{3}{8} \epsilon_pf_{p\gamma}(\epsilon_p)L_{\epsilon_p}
,\end{equation}
where $\epsilon_pL_{\epsilon_p}$ is the  differential proton luminosity, and the neutrinos emerge with an energy approximately given by $\epsilon_{\nu} \sim \epsilon_p/20$, $f_{p\gamma}$ intuitively represents the production efficiency of neutrinos.
Figure \ref{f6} shows the neutrino production efficiency for GB6 J2113+1121 and NVSS J171822+423948 in different soft photon fields. It can be seen that the external photon field from the BLR can significantly enhance the neutrino luminosity. In other words, in the absence of BLR photons for $p\gamma$ interaction, achieving substantial neutrino luminosity would require the proton luminosity of GB6 J2113+1121 and NVSS J171822+423948 to far exceed their respective Eddington luminosities by multiple orders of magnitude.
From an energy budget perspective, the radiation regions of GB6 J2113+1121 and NVSS J171822+423948 are most likely located within or near the BLR.

\begin{figure*}
    \centering
    \includegraphics[scale=0.5]{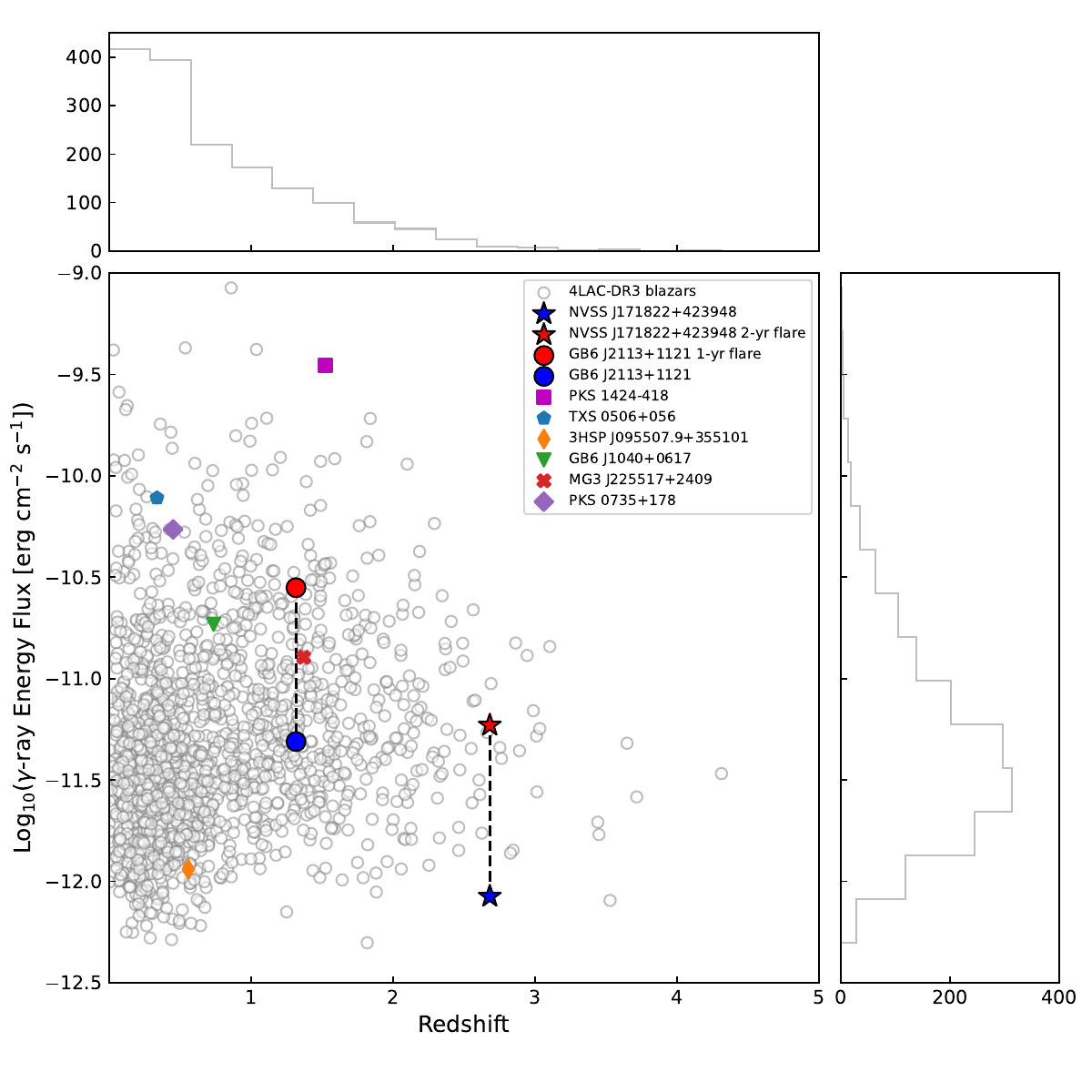}
    \includegraphics[scale=0.5]{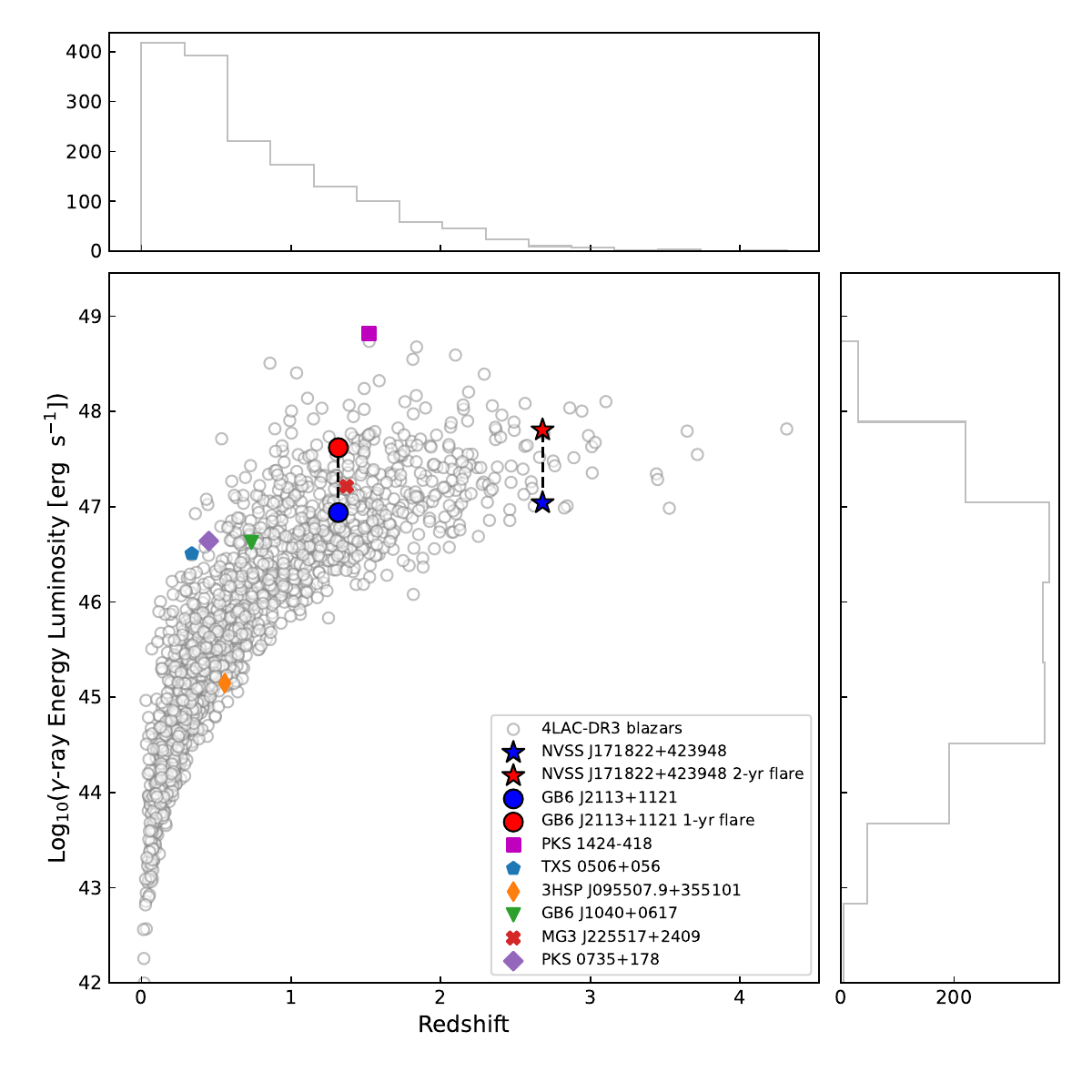}
    \caption{A comparison between the known neutrino-emitting candidates and 4LAC-DR3 blazars, with the known neutrino-emitting candidates highlighted in color. The energy ranges corresponding to the energy flux and luminosity are between 100 MeV and 100 GeV. In the right panel, the luminosities have been corrected for K-correction.}
    \label{f7}
\end{figure*}

\begin{figure*}
    \centering
    \includegraphics[scale=0.8]{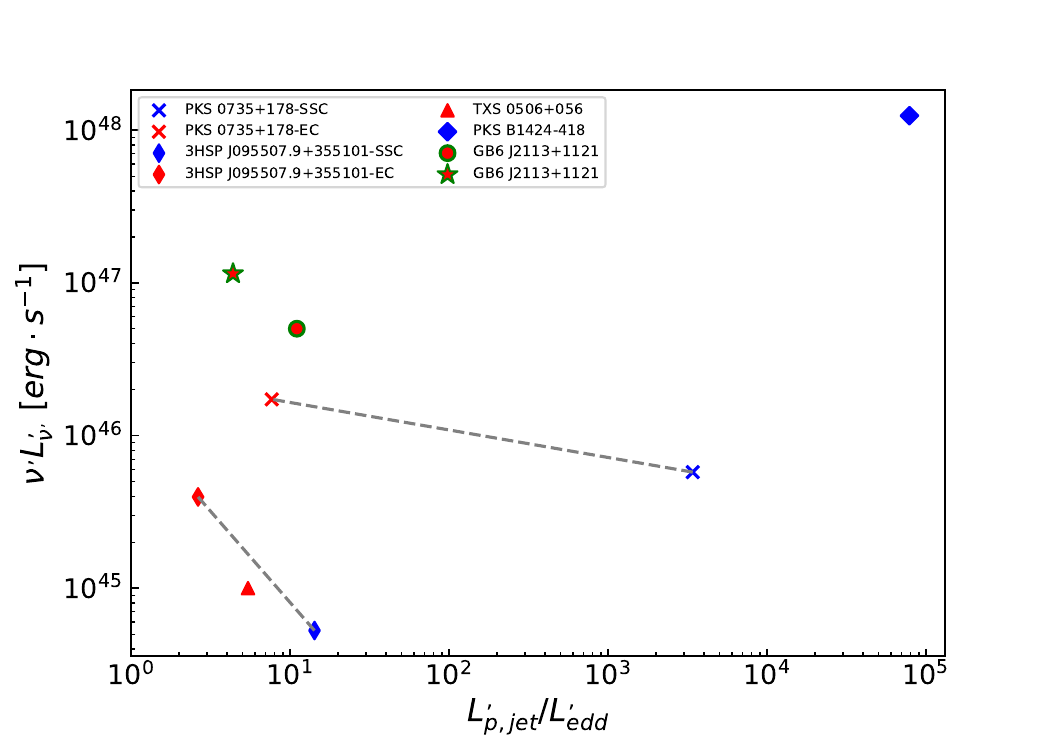}
    \caption{The comparison of the jet power (In units of their respective Eddington luminosities) of different neutrino candidates with the peak luminosity of all flavor neutrinos. The red points indicate that the soft photon field in the $p\gamma$ process comes from an external photon field, while the blue points indicate that the soft photon field in the $p\gamma$ process comes solely from the synchrotron radiation of primary electrons. For TXS 0506+056, the the neutrino peak luminosity and jet power comes from \cite{2018ApJ...864...84K}, and it is assumed that the mass of its supermassive black hole (SMBH) is a typical value of $1\times 10^9 M_\odot$ \citep{2019ApJ...886...23X}. For PKS 0735+178, the neutrino peak luminosity, jet power, and the SMBH mass all come from \cite{2023MNRAS.519.1396S}.(see their Hybrid and Hybrid-ext models). For 3HSP J095507.9+355101, the neutrino peak luminosity and jet power come from \cite{2020ApJ...899..113P}(see their Model A(B = 15 G) and HEP Scenario), while the SMBH mass comes from \cite{2020A&A...640L...4G}. The neutrino peak luminosity and jet power for PKS B1424-418 come from \cite{2017ApJ...843..109G}(see their joint best-fit  for SED and neutrino in the burst state). For GB6 J2113+1121 and NVSS J171822+423948, the SMBH masses come from \cite{2022ApJ...932L..25L} and \cite{2024ApJ...965L...2J}, respectively.}
    \label{f8}
\end{figure*}

It is particularly intriguing to compare GB6 J2113+1121 and  NVSS J171822+423948 with other flare blazars linked to neutrino events. These flare blazars include TXS 0506+056 (ISP; \citep{1995ApJ...444..567P}), GB6 J1040+0617 (LSP; \citep{2019ApJ...880..103G}), MG3 J225517+2409, PKS 0735+178 (ISP; \citep{2020ApJ...893..162F}; \citep{2023MNRAS.519.1396S}), 3HSP J095507.9+355101 (HSP; \citep{2020A&A...640L...4G}), as well as the high-luminosity FSRQ PKS B1424-418 (LSP; \citep{2016NatPh..12..807K}). This suggests that blazars with different $\nu_{peak}^S$ may all contribute to IceCube neutrinos. Furthermore, \citep{2020MNRAS.497..865G} have demonstrated a 3.2 $\sigma$ level association between IceCube tracks and the positions of ISPs and HSPs. Among the flaring blazars linked to neutrino events, GB6 J1040+0617, MG3 J225517+2409, and 3HSP J095507.9+355101 are BL Lacs with intrinsically weak emission lines or lacking emission lines altogether. GB6 J2113+1121, NVSS J171822+423948, and PKS B1424-418 are FSRQs with broad emission lines.  TXS 0506+056 and PKS 0735+178 are considered masquerading BL Lacs, meaning they are actually FSRQs with emission lines overshadowed by a very bright, Doppler-boosted jet \citep[e.g.,][]{2019MNRAS.484L.104P, 2023MNRAS.519.1396S}. The latest incremental version of the fourth catalog of AGNs detected by Fermi-LAT (4LAC-DR3; \citep{2022ApJS..263...24A}) includes approximately 650 FSRQs and about 1350 BL Lacs. However, considering that among the neutrino candidates, there are 5 FSRQs (e.g., GB6 J2113+1121, NVSS J171822+423948, PKS B1424-418, TXS 0506+056, and PKS 0735+178) and 3 BL Lacs (e.g., 3HSP J095507.9+355101, GB6 J1040+0617, and MG3 J225517+2409), this suggests that, from an observational perspective, flaring FSRQs are more likely to be associated with IceCube neutrinos. As shown in the left panel of Figure \ref{f7}, four sources have a redshift lower than 1 (TXS 0506+056, PKS 0735+178, 3HSP J095507.9+355101, GB6 J1040+0617), while four sources have a redshift higher than 1 (GB6 J2113+1121, MG3 J225517+2409, PKS B1424-418, NVSS J171822+423948). This distribution is roughly consistent with the redshift distribution of 4LAC-DR3 blazars, where the number of blazars with redshift lower than 1 is approximately twice that of those with redshift greater than 1. Among all neutrino candidate sources, PKS B1424-418, TXS 0506+056, and PKS 0735+178 are the three sources with the highest $\gamma-$ray flux. They are also high $\gamma-$ray flux sources in 4LAC-DR3, with PKS B1424-418 being one of the brightest $\gamma-$ray blazars in the entire sky. 3HSP J095507.9+355101, GB6 J2113+1121 and NVSS J171822+423948 have the lowest $\gamma-$ray flux, with NVSS J171822+423948 remaining notably faint even among the high-redshift blazars ($z \geq 2.5$) in 4LAC-DR3. This seems to suggest that there is no positive correlation between the neutrinos detected by IceCube and the $\gamma-$ray brightness. However, GB6 J2113+1121 and NVSS J171822+423948 have remained below the Fermi-LAT detection threshold for a long period, and around the time of the neutrino arrival, their respective $\gamma$-ray fluxes significantly increased to relatively high levels (see the red points in the left panel of Figure \ref{f7}). It can be seen that sources with higher redshift tend to have higher $\gamma-$ray luminosity (see the right panel of Figure \ref{f7}), suggesting that jets in the early universe were generally more powerful. Although GB6 J2113+1121 and NVSS J171822+423948 have low $\gamma$-ray flux, their $\gamma$-ray luminosity is high considering their high redshift, several times higher than that of TXS 0505+056. It can also be seen that, except for 3HSP J095507.9+355101 (an X-ray flaring-selected neutrino candidate that did not exhibit a $\gamma-$ray flare at the time of the neutrino arrival), all other neutrino candidates have relatively high $\gamma-$ray luminosity, with $\gamma-$ray luminosities all exceeding $3 \times 10^{46}$ erg s$^{-1}$. This suggests that neutrino production may be linked to the intrinsic $\gamma$-ray luminosity, although many of the most luminous blazars have not been detected as neutrino sources.   Figure \ref{f8} shows the relationship between jet power and peak neutrino luminosity for different neutrino candidates modeled using the One-zone Leptohadronic scenario \footnote{We have not considered
the proton-synchrotron model, which primarily produces the high energy component through proton synchrotron radiation and requires very high proton energy, resulting in a neutrino peak luminosity greater
than 100 PeV, exceeding IceCube’s detection range. Additionally, we did not plot GB6 J1040+0617 and MG3
J225517+2409 because there are no available one-zone leptohadronic model results for these two sources.}. It can be seen that for all the sources, their jet power exceeds the Eddington luminosity ($L_{p,jet}^{'}/L_{edd}^{'} \geq 1$). The fact that $L_{p,jet}^{'}/L_{edd}^{'} \geq 1$ implies that, unless the possibility of super-Eddington accretion is assumed, the jet power cannot be maintained at this level for a long time. However, blazars remain in a relatively quiescent state for most of their lifetime, and the $\gamma$-ray flares associated with neutrino events last only for a short period within their lifetime, making short-term super-Eddington jet power plausible. Therefore, a jet with super-Eddington luminosity may be an important condition for the production of IceCube neutrinos. It can also be seen that, when the jet power does not significantly exceed the Eddington luminosity ($L_{p,jet}^{'}/L_{edd}^{'} \leq 10$), GB6 J2113+1121 and NVSS J171822+423948 have the highest peak neutrino luminosities, reaching up to $\sim$ $5\times 10^{46}$ erg s$^{-1}$ and $\sim$ $1\times 10^{47}$ erg s$^{-1}$ respectively. Additionally, for PKS 0735+178 and 3HSP J095507.9+355101, the peak neutrino luminosity shows a significant increase when considering an external photon field from the hidden BLR region, and the required jet power is also significantly reduced. The FSRQ PKS B1424-418 has the highest neutrino luminosity, its peak neutrino luminosity is $\sim$ $1\times 10^{48}$ erg s$^{-1}$, while the required jet power exceeds the Eddington luminosity by about five orders of magnitude. This is because, despite the strong BLR emission of FSRQ PKS B1424-418, \cite{2017ApJ...843..109G} estimated a rather low magnetic field strength, suggesting that the radiation region is located outside both the BLR and the dusty torus. Therefore, they did not account for the external photon field from the BLR in their modeling, which explains the need for such a high jet power.

An interesting example is that the neutrino event IC-190730A was found to be spatially coincident, but not temporally, with the bright $\gamma$-ray blazar PKS 1502+106.
At the time of the neutrino alert, PKS 1502+106 was in a quiet state in UV/optical/X-ray/$\gamma$-ray flux. \cite{2021JCAP...10..082O} suggested that the neutrino might be linked to PKS 1502+106's long-term quiet emission. 
Their modeling results indicate that, when the jet power is below the Eddington luminosity, the multi-wavelength and neutrino emission of PKS 1502+106 originates beyond the BLR and inside the dust torus, which is most consistent with the observations.
PKS 1502+106 is similar to PKS B1424-418 in that both exhibit strong BLR radiation, with their emission regions located outside the BLR. This suggests that some other FSRQs might also have similar characteristics, namely that their radiation regions are located outside the BLR. Additionally, there is evidence that some FSRQs seem to have emitting regions beyond the BLR \citep[e.g.,][]{2018MNRAS.477.4749C,2019ApJ...877...39M, 2021MNRAS.500.5297A}. The absence of the BLR photon field as the target for the $p\gamma$ process, resulting in ineffective production of TeV-PeV neutrinos, might be one reason why IceCube has not detected strong stacking \citep[e.g.,][]{2017ApJ...835...45A,2017arXiv171001179I} or clustering \citep{2015APh....66...39A,2016PhRvD..94j3006M} signals of neutrinos from FSRQs, as well as imposing constraints on the diffuse neutrino flux \citep[e.g.,][]{2018PhRvD..98f2003A, 2020ApJ...890...25Y}. Although the radiation regions of GB6 J2113+1121 and NVSS J171822+423948 are within the BLR, their large distances from us, particularly NVSS J171822+423948 with a redshift of 2.7, result in their $\gamma-$rays and neutrino fluxes being too faint. The large distance may be one of the reasons why high-redshift FSRQs are not considered the favored primary source of the observed IceCube neutrinos, although they are likely to be effective neutrino emission sources.

 In the 4LAC-DR3, there are a total of $\sim$ 3700 $\gamma$-ray blazars, of which $\sim$ 560 have redshifts greater than 1, constituting a significant portion of the total sample. Only 38 blazars have redshifts greater than 2.5, indicating that  NVSS J171822+423948 is a rather rare high-redshift blazar, which is crucial for understanding the formation, evolution, and radiation mechanisms of early cosmic jets \citep{2017ApJ...837L...5A, 2018ApJ...853..159L,2018ApJ...865L..17L,2019ApJ...879L...9L}. The spatial and temporal coincidence of the neutrino IceCube-201221A with NVSS J171822+423948 suggests that neutrinos may originate from such high-redshift blazars, implying that in the early universe, blazar jets are effective cosmic ray and neutrino emitters, with hadronic processes potentially playing a significant role in their radiation. Although FSRQs are disfavored as the dominant origin of the observed IceCube neutrinos, their inherent high luminosity, coupled with the strong photon fields associated with BLRs, makes them excellent candidates for neutrino production.  Therefore, individual flaring FSRQs are still promising and offer hopeful opportunities for the identification of neutrino emitters. Aside from GB6 J2113+1121 and NVSS J171822+423948, we anticipate that IceCube will detect more neutrinos from FSRQs with redshifts greater than 1 in the future.  A notable example is the FSRQ PKS 0215+015, located at z = 1.715 \citep{1987AJ.....93..529F}, which was experiencing a significant $\gamma$-ray flare at the time of the spatially coincident neutrino event IceCube-220225A \citep{2022GCN.31653....1G}.

\section{Summary} \label{sec:5}

In this work, due to the temporal and spatial correlations between the FSRQ GB6 J2113+1121 and the neutrino IceCube-191001A, as well as between NVSS J171822+423948 and the neutrino IceCube-201221A, we applied the one-zone leptohadronic scenario to model the multi-wavelength electromagnetic and neutrino emissions of both FSRQs, taking into account an external photon field from the BLR as the target photon field for  neutrino production. We found that the synchrotron emission from pair cascades induced by hadronic processes can make a dominant contribution to the X-ray and GeV $\gamma-$ray emission, while at other wavelengths, the emission from primary electrons is dominant. Our modeling results for both FSRQs have yielded a low neutrino flux level.  For GB6 J2113+1121, the maximum muon neutrino flux is $\sim 1\times10^{-12}$ erg cm$^{-2}$ s$^{-1}$, peaking at $\sim$ 3 PeV. During the 1-year $\gamma$-ray flaring state, the probability of detecting a single muon neutrino is 2$\%$. For  NVSS J171822+423948, the maximum muon neutrino flux is $\sim 6\times10^{-13}$ erg cm$^{-2}$ s$^{-1}$, peaking at $\sim$ 2 PeV. During the two-year $\gamma$-ray flaring state, the probability of detecting a single muon neutrino is approximately 0.8$\%$. We did not consider the SSC scenario, where synchrotron photons from primary electrons are the only targets for $p\gamma$ interactions, as neutrinos cannot be effectively produced in this scenario. We also did not consider the proton-synchrotron scenario, where proton synchrotron radiation accounts for the high-energy hump in the SED. This requires very high proton energies, typically exceeding EeV, which would result in a neutrino spectrum peaking beyond 100 PeV, leading to a highly suppressed sub-PeV neutrino flux.

Although FSRQs contribute little to the diffuse neutrino background, it is possible for a single source to contribute to IceCube neutrinos, depending on whether its emission region is close to the BLR. We found that although the photon field in the BLR can significantly enhance neutrino production efficiency, it also leads to significant absorption of high-energy $\gamma$-rays (above tens of GeV) produced by cascade emissions associated with neutrinos, thereby complicating the relationship between neutrinos and $\gamma$-rays. Nonetheless, cascade emission inevitably makes a significant contribution in the X-ray regime, which allows for better constraints on hadronic processes through X-ray observations. We also found that, unlike GB6 J2113+1121 and  NVSS J171822+423948, which have their emission regions within the BLR, the other two FSRQs associated with neutrinos, PKS B1424-418 and PKS 1502+106, have their emission regions outside the BLR. Additionally, there is evidence suggesting that a significant portion of the $\gamma$-ray emission region of FSRQs lies outside the BLR. Considering that both neutrinos and $\gamma-$rays originate from the same emission region, and that this region is located outside the BLR, the efficiency of neutrino production is relatively low. This could be one of the reasons why FSRQs are not a major source of IceCube neutrinos. On the other hand, although some FSRQs have their emission regions within or near the BLR, the neutrino flux from them, such as GB6 J2113+1121 and NVSS J171822+423948, remains too low due to their great distance from Earth. Moreover, there are few such sources, which may also contribute to why FSRQs are not considered a major source of IceCube neutrinos.  

However, for  NVSS J171822+423948, there is no available X-ray data to constrain its hadronic component. We hope to conduct future X-ray observations of this source, as they are crucial for understanding the radiation, evolution, and other processes associated with high-redshift blazars.

We acknowledge the use of publicly available Python libraries, including {ASTROPY} \citep{2013A&A...558A..33A}, {MATPLOTLIB} \citep{hunter2007matplotlib}, {PANDAS} \citep{reback2020pandas}, {NUMPY} and {SCIPY} \citep{2020NatMe..17..261V}, which have greatly facilitated our analysis and modeling efforts. We utilized publicly released data from the IceCube Neutrino Observatory and received assistance from the NASA/IPAC Extragalactic Database and the SIMBAD database.

This work was partially supported by the National Natural Science Foundation of China (NSFC) under grants U2031120, 12203043, 12473049, 12073080, 11933010, and 11921003.
This work was also supported in part by the Special Natural
Science Fund of Guizhou University (grant No. 201911A) and
the First-class Physics Promotion Programme (2019) of
Guizhou University.

\appendix   

\section{Appendix: multi-band light curve}
\begin{figure*}
    \centering
    \includegraphics[scale=0.5]
    {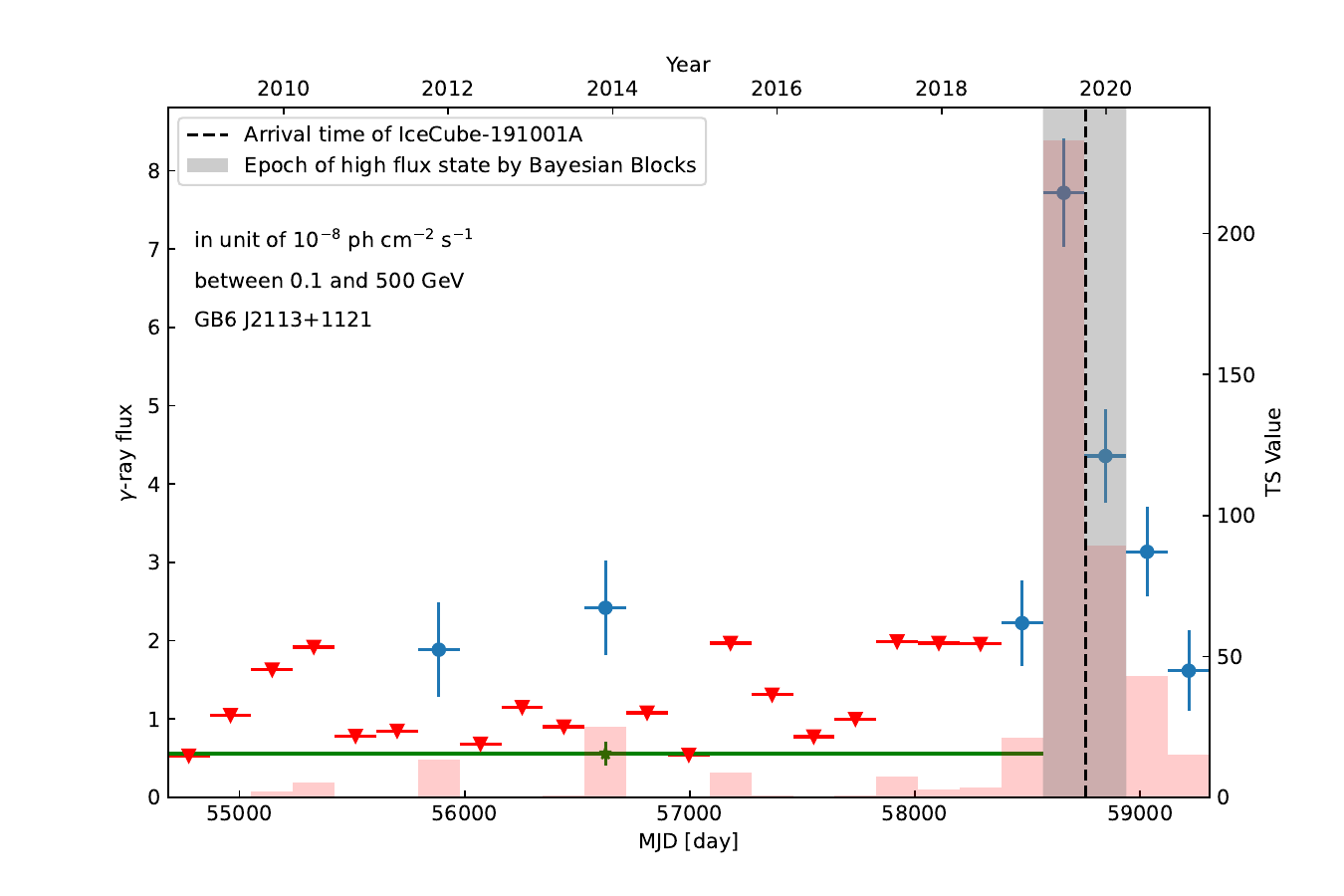}
    \caption{Half-year time bin $\gamma$-ray light curves of GB6 J2113+1121. Blue circles and red triangles represent flux estimations and upper limits, respectively, and the TS values corresponding to each time bin are also shown. The gray shaded area represents an epoch of high $\gamma$-ray flux state obtained by Bayesian blocks. The green horizontal line indicates the average flux of the source during its quiescent state over a long period. The black dashed line marks the arrival time of the neutrino IceCube-191001A.}
    \label{fA.1}
\end{figure*}

\begin{figure*}
    \centering
    \includegraphics[scale=0.5]
    {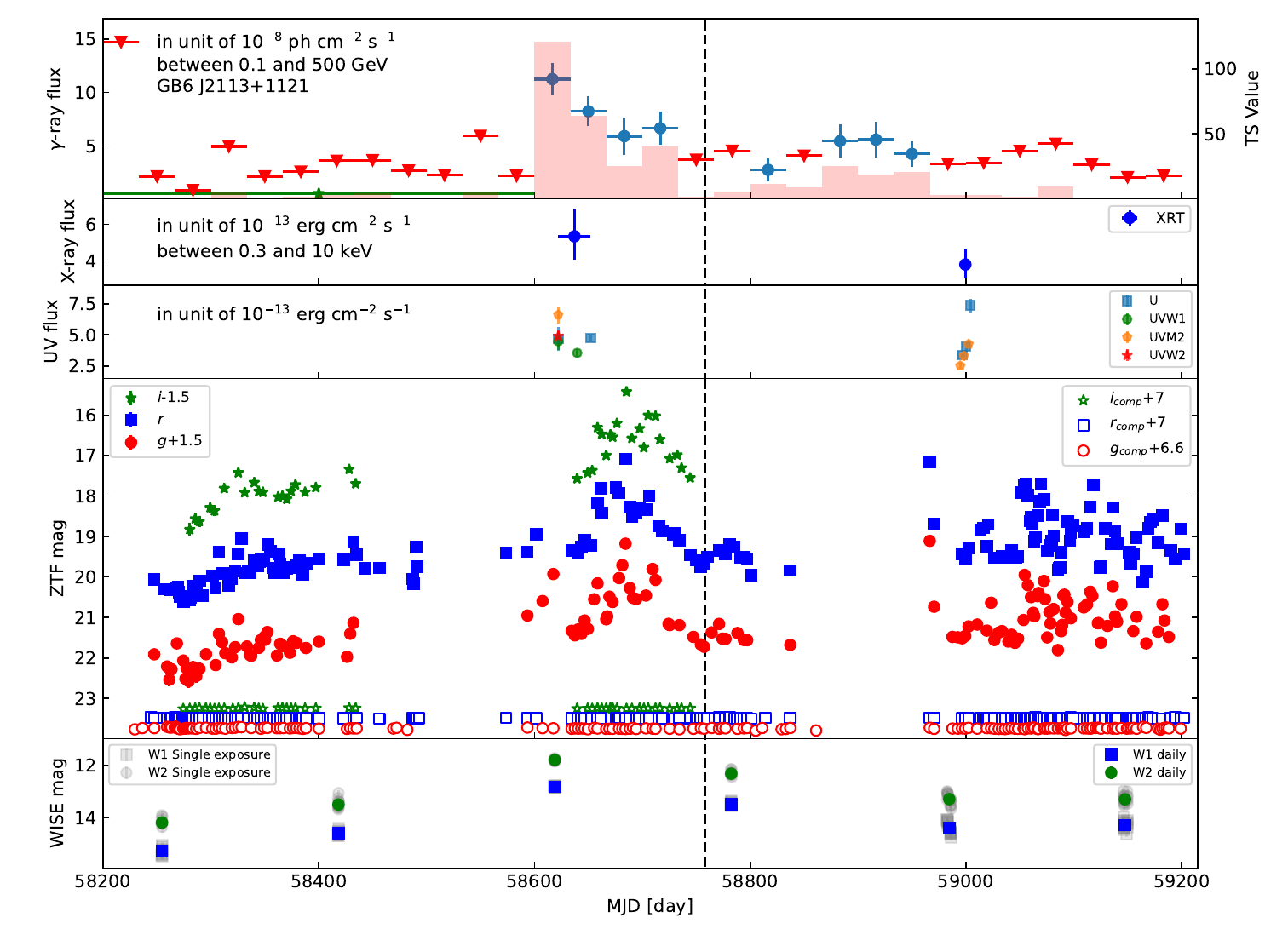}
    \caption{Multiwavelength light curves of GB6 J2113+1121. In the monthly $\gamma-$ray light-curve panel, all lines and points are the same as those in Figure \ref{fA.1}. For the ZTF light curves, the solid markers represent the magnitudes of the target, while the hollow ones correspond to the average magnitudes of the comparison stars in the same field. The black dashed vertical line across all panels marks the arrival time of the neutrino.}
    \label{fA.2}
\end{figure*}

\begin{figure*}
    \centering
    \includegraphics[scale=0.5]
    {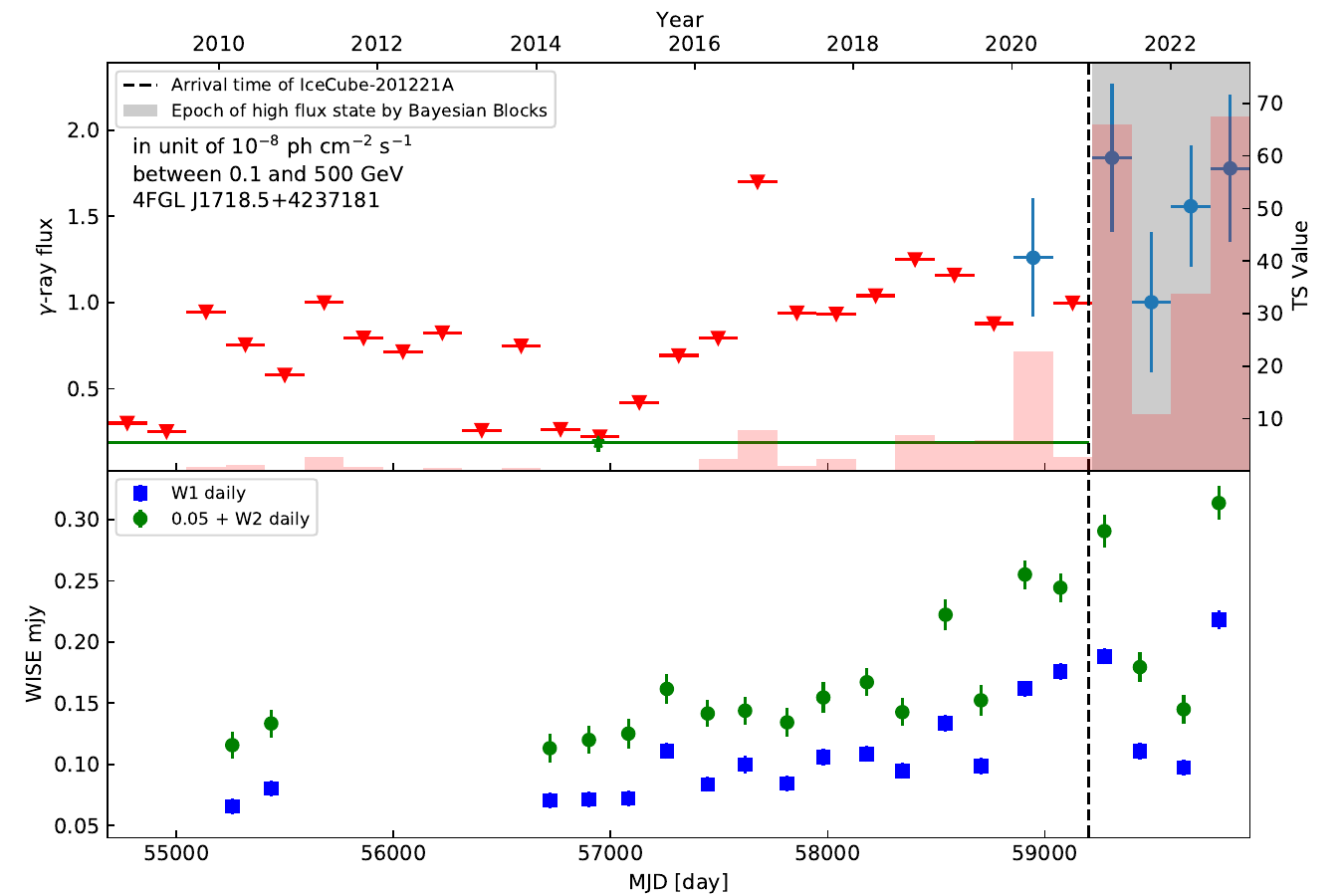}
    \caption{The $\gamma$-ray (half-year time bin) and infrared light curves of NVSS J171822+423948. In the former, blue circles and red triangles correspond to flux estimations and upper limits, with red bars representing TS values. The gray shaded area indicates a period of high $\gamma$-ray flux state obtained through the Bayesian blocks method. The green horizontal line represents the long-term average flux of NVSS J171822+423948 in its quiescent state. The black dashed vertical line across the light curves marks the arrival time of IceCube-201221A.}
    \label{fA.3}
\end{figure*}

\begin{figure*}
    \centering
    \includegraphics[scale=0.5]
    {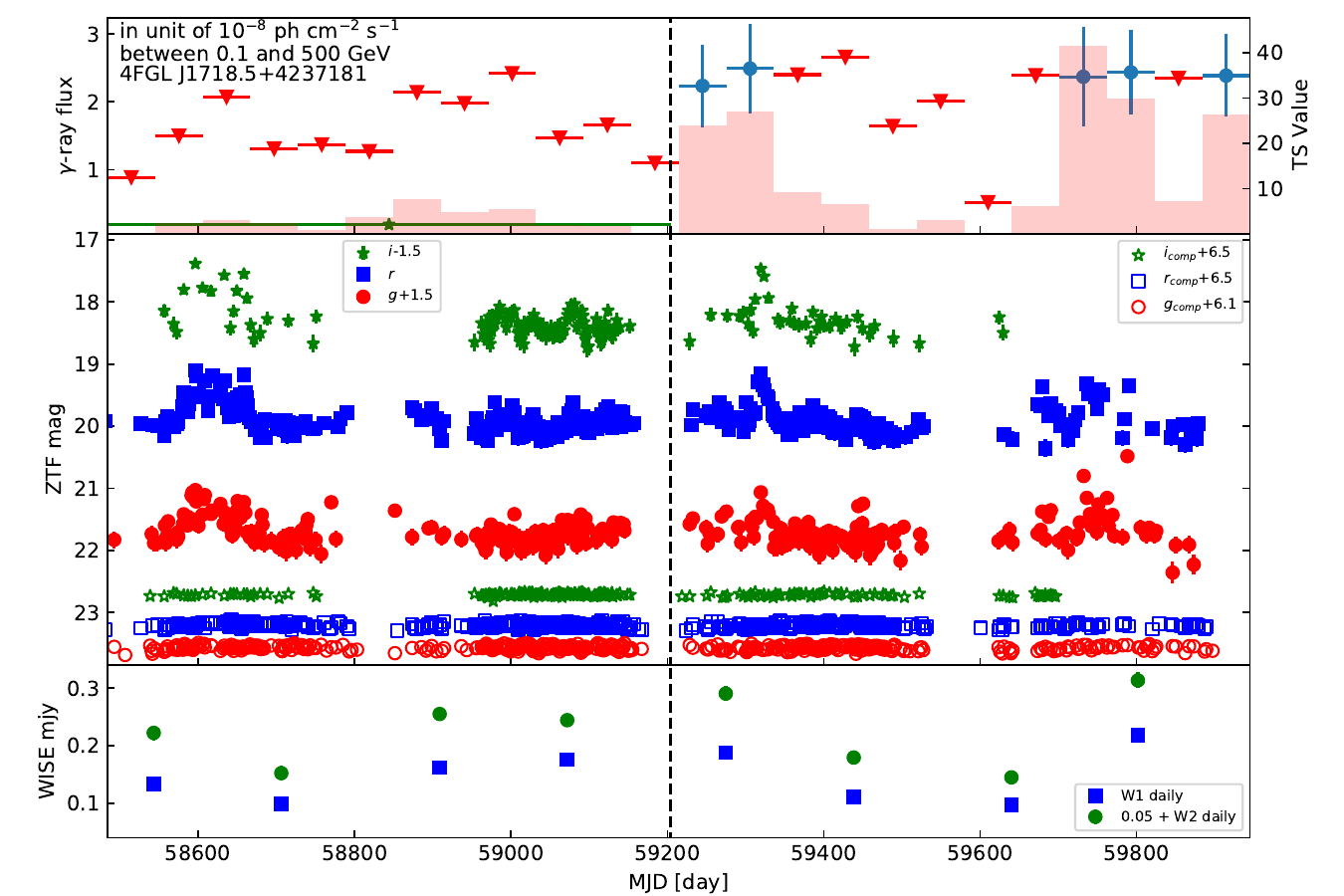}
    \caption{Zoomed-in multiwavelength light curves. Upper panel: 2-month time-bin $\gamma$-ray light curves of 4FGL J1718.5+4237, with all lines and points identical to those in the upper panel of Figure \ref{fA.3}. Middle panel: ZTF light curves of NVSS J171822+423948, where solid markers represent the magnitudes of the target, while hollow ones correspond to the average values of the comparison stars in the same field. Bottom panel: WISE light curves of NVSS J171822+423948. A black vertical dashed line across all panels marks the arrival time of the neutrino IceCube-201221A.}
    \label{fA.4}
\end{figure*}

To better understand the relationship between neutrino and electromagnetic observations, we present the multi-wavelength light curves of GB6 J2113+1121 and NVSS J171822+423948 in this section. Figure \ref{fA.1} shows the half-year time bin $\gamma$-ray light curves of GB6 J2113+1121, where intense $\gamma$-ray activity can be observed approximately half a year before the arrival of the neutrino (grey dashed line). This flaring activity lasted for about one year (black shaded region). Figure \ref{fA.2} presents a zoomed-in view of the multi-wavelength light curves of GB6 J2113+1121. Figure \ref{fA.3} displays the half-year time bin $\gamma$-ray light curves and infrared light curves of NVSS J171822+423948, indicating that after the arrival of the neutrino IceCube-201221A, the source underwent a $\gamma$-ray flare lasting for about two years, accompanied by an increase in infrared flux. Figure \ref{fA.4} presents a zoomed-in view of the multi-wavelength light curves of NVSS J171822+423948.

\bibliography{refs}
\bibliographystyle{aasjournal}

\begin{table}[ht]
    \centering
    \caption{Parameter values used for the SED models}
    \begin{tabular}{lcc}
        \toprule
        Parameter & GB6 J2113+1121 & NVSS J171822+423948 \\ 
        \midrule
        $\delta$    & 30   & 20   \\ 
        $R$ (cm)    & 1$\times$10$^{16}$   & 3$\times$10$^{16}$   \\ 
        $B$ (G)    & 6   & 2.5   \\ 
        $r_{in}^{'}$ (pc)    & 0.1   & 0.2 \\
        $\gamma_{e,min}$    & 50   & 50 \\
        $\gamma_{e,break}$    & 1$\times$10$^3$   & 2$\times$10$^3$ \\
        $\gamma_{e,max}$       & 7.8$\times$10$^{3}$   & 6.5$\times$10$^{4}$\\
        $n_{e,1}$    & 1.2   & 1.6   \\
        $n_{e,2}$    & 3.9   & 3.9   \\
        $\gamma_{p,min}$    & 1   & 1   \\ 
        $\gamma_{p,max}$    & 2$\times$10$^7$   & 2.5$\times$10$^7$   \\
        $n_{p}$    & 2   & 2   \\
        \\
        $L_{e,inj}$ (erg s$^{-1}$)    & 9.8$\times$10$^{41}$   & 7.8$\times$10$^{42}$   \\
        $L_{p,inj}$ (erg s$^{-1}$)   & 1.4$\times$10$^{44}$   & 8.9$\times$10$^{43}$   \\
        $L_{B,jet}^{'}$  (erg s$^{-1}$)  & 1.2$\times$10$^{46}$   & 8.4$\times$10$^{45}$   \\
        $L_{e,k}^{'}$ (erg s$^{-1}$)   & 2.0$\times$10$^{43}$   & 4.4$\times$10$^{43}$   \\
        $L_{p,k}^{'}$ (erg s$^{-1}$)   & 9.8$\times$10$^{46}$   & 2.6$\times$10$^{47}$   \\
        \bottomrule
    \end{tabular}
    \label{t1}
\end{table}

\end{document}